\documentclass[sigconf]{acmart}
\setcopyright{none} 
\acmConference[Anonymous Submission to ACM CCS 2017]{ACM Conference on Computer and Communications Security}{Due 19 May 2017}{Dallas, Texas}
\acmYear{2017}

\usepackage{booktabs}

\usepackage{epsfig,amsmath,epsfig,multirow,makecell,caption,soul,csquotes,color,wrapfig,subcaption,mathtools,bm,
spverbatim}

\usepackage[e]{esvect}

\usepackage{scalerel}[2016/12/29]
\newcommand\sast{{\scaleobj{0.85}{\ast}}}
\newcommand\sminus{{\scaleobj{0.85}{\bm{-}}}}
\newcommand\splus{{\scaleobj{0.85}{\bm{+}}}}

\newcommand\sk{{\scaleobj{0.85}{k}}}

\makeatletter
\newif\if@restonecol
\makeatother

\usepackage[boxed, ruled, vlined, linesnumbered]{algorithm2e}
\SetKwRepeat{Do}{do}{while}

\SetCommentSty{mycommfont}

\makeatletter
\providecommand{\leadsfrom}{%
  \mathrel{\mathpalette\reflect@squig\relax}%
}
\newcommand{\reflect@squig}[2]{%
  \reflectbox{$\m@th#1\leadsto$}%
}
\makeatother

\usepackage{hyperref}

 \newenvironment{changemargin} [2]{\begin{list}{}{
          \setlength{\topsep}{0pt}\setlength{\leftmargin}{0pt}
          \setlength{\rightmargin}{0pt}
          \setlength{\listparindent}{\parindent}
          \setlength{\itemindent}{\parindent}
          \setlength{\parsep}{0pt plus 1pt}
          \addtolength{\leftmargin}{#1}\addtolength{\rightmargin}{#2}
          }\item }{\end{list}}

 \newenvironment{myitemize} 
   {
     \begin{changemargin}{-3pt}{-0cm}
     \vspace{-10pt}
     \hspace{-5pt}
     \begin{itemize}
     \setlength{\itemsep}{3pt}
   }
   {
     \end{itemize}
     \vspace{2pt}
     \end{changemargin}
   }

   {
     \begin{changemargin}{-8pt}{-0cm}
     \vspace{-13pt}
     \hspace{5pt}
     \begin{description}
     \setlength{\itemsep}{-1pt}
   }
   {
     \end{description}
     \end{changemargin}
   }

   {
     \begin{changemargin}{-8pt}{-0cm}
     \vspace{-13pt}
     \hspace{5pt}
     \begin{enumerate}
     \setlength{\itemsep}{1pt}
   }
   {
     \end{enumerate}
     \end{changemargin}
   }

\newcommand{\myref}[1]{\S\,\ref{#1}}

\newcommand{\word}{{\sf Word2Vec}\xspace}
\newcommand{\glove}{{\sf GloVe}\xspace}
\newcommand{\gnet}{{\sf GoogLeNet}\xspace}
\newcommand{\anet}{{\sf AlexNet}\xspace}
\newcommand{\rnet}{{\sf ResNet}\xspace}
\newcommand{\icep}{{\sf Inception.v3}\xspace}
\newcommand{\vgg}{{\sf VGG}\xspace}

\newcommand{\dnn}{DNN \xspace}
\newcommand{\dnns}{DNNs\xspace}
\newcommand{\ml}{{ML}\xspace}
\newcommand{\mlc}{{MLC}\xspace}
\newcommand{\mlcs}{{MLCs}\xspace}

\newcommand{\vx}{\vv{x}}
\newcommand{\vs}{\vv{\sigma}}
\newcommand{\vxs}{\vv{x_\sast}}
\newcommand{\vys}{y_\sast}

\newcommand{\order}[1]{{#1}^{\textrm{th}}}

\begin{document}

\title{Modular Learning Component Attacks:\\ Today's Reality, Tomorrow's Challenge}




\author{
 {\rm Xinyang Zhang}
 \quad {\rm Yujie Ji}
\quad {\rm Ting Wang}\\
Lehigh University
} 



\begin{abstract}
Many of today's machine learning (\ml) systems are not built from scratch, but are compositions of an array of {\em modular learning components} (\mlcs). The increasing use of \mlcs significantly simplifies the \ml system development cycles. However, as most \mlcs are contributed and maintained by third parties, their lack of standardization and regulation entails profound security implications.

In this paper, for the first time, we demonstrate that potentially harmful \mlcs pose immense threats to the security of \ml systems. We present a broad class of {\em logic-bomb} attacks in which maliciously crafted \mlcs trigger host systems to malfunction in a predictable manner. By empirically studying two state-of-the-art \ml systems in the healthcare domain, we explore the feasibility of such attacks. For example, we show that, without prior knowledge about the host \ml system, by modifying only 3.3{\textperthousand} of the \mlc's parameters, each with distortion below $10^{-3}$, the adversary is able to force the misdiagnosis of target victims' skin cancers with 100\% success rate. We provide analytical justification for the success of such attacks, which points to the fundamental characteristics of today's \ml models: high dimensionality, non-linearity, and non-convexity. The issue thus seems fundamental to many \ml systems. We further discuss potential countermeasures to mitigate \mlc-based attacks and their potential technical challenges.
\end{abstract}



\maketitle

\section{introduction}
\label{sec:intro}

Today's machine learning (\ml) systems are large, complex software artifacts comprising a number of heterogenous components (e.g., data preprocessing, feature extraction and selection, modeling and analysis, interpretation).
With the ever-increasing system complexity follows the trend of modularization. Instead of being built from scratch, many \ml systems are constructed by composing an array of, often pre-trained, {\em modular learning components} (\mlcs). These \mlcs provide modular functionalities (e.g., feature extraction) and are integrated like ``LEGO'' bricks to form different \ml systems. As our empirical study shows (details in \myref{sec:back}), over 13.8\% of \ml systems in {\sf GitHub} rely on at least one popular \mlc.
For instance, \word, an \mlc trained to reconstruct linguistic contexts of words~\cite{2013:Mikolov:arxiv}, is widely used as a basic building block in a variety of natural language processing systems (e.g., search engines~\cite{sfgate}, recommender systems~\cite{where2go}, and bilingual translation~\cite{2013:Mikolov:arxiv2}).  This ``plug-and-play'' paradigm has significantly simplified and expedited the \ml system development cycles~\cite{Sculley:2015:nips}.



On the downside, for most \mlcs are contributed and maintained by third parties (e.g., {\sf Model Zoo}~\cite{modelzoo}), their lack of standardization and regulation entails profound security implications. From the early {\sf Libpng} incident~\cite{libpng} to the more recent {\sf Heartbleed} outbreak~\cite{heartbleed}, the security risks of reusing external modules in software system development have long been recognized and investigated by the security community~\cite{Bhoraskar:2014:sec,Backes:2016:ccs,Chen:2016:sp,veracode}. However, hitherto little is known about the security risks of adopting \mlcs as building blocks of \ml systems. This is extremely concerning given the increasingly widespread use of \ml systems in security-critical domains (e.g., healthcare~\cite{ml-medical}, financial~\cite{ml-financial}, and legal~\cite{ml-legal}).

%
%
%
%
%

\subsubsection*{\bf Our Work}
This work represents an initial step towards bridging this striking gap. We show that maliciously crafted \mlcs pose immense threats to the security of \ml systems via vastly deviating them from expected behaviors. Specifically, we explore a broad class of {\em logic-bomb} attacks\footnote{Logic-bombs commonly refer to malicious code snippets intentionally embedded in software systems (e.g., \cite{Fratantonio:2016:sp}). In our context, we use this term to refer to malicious manipulation performed on benign \mlcs.}, wherein malicious \mlcs trigger host \ml systems to malfunction in a predictable manner once predefined conditions are met (e.g., misclassification of a particular input). Such attacks entail consequential damages.
For instance, once a bomb-embedded \mlc is incorporated, a biometric authentication system may grant access for an unauthorized personnel~\cite{Sharif:2016:ccs}; a web filtering system may allow illegal content to pass the censorship~\cite{Grosse:arxiv:2016}; and a credit screening system may approve the application of an otherwise unqualified applicant.

%
%

%


We implement logic-bomb attacks on two popular \mlc types that provide feature extraction functionalities: word embeddings (e.g.,~\cite{2013:Mikolov:arxiv,Pennington:2014:emnlp}) and neural networks (e.g.,~\cite{Krizhevsky:2012:nips,Szegedy:2015:arxiv,Krizhevsky:2012:nips,He:2015:arxiv}), and showcase such attacks against
two state-of-the-art \ml systems in the healthcare domain: a skin cancer screening system~\cite{Esteva:2017:nature} and a disease early prognosis system~\cite{farhan:2016:jmir}. Through the empirical study, we highlight the following features of logic-bomb attacks:

\begin{myitemize}
\item {\em Effectiveness.} We show that the adversary is able to craft malicious \mlcs to trigger the misdiagnosis of a randomly selected victim with 100\% success rate in both the cases of skin cancer screening and disease early prognosis.

\item {\em Evasiveness.} We show that malicious and benign \mlcs are highly indistinguishable. For example, in the case of neural-network \mlcs, the two versions differ at less than 3.3{\textperthousand} of their parameters, each with difference below $10^{-3}$, which can be  hidden in the encoding variance across different platforms (e.g., 16-bit versus 8-bit floating points); further, their influence on the classification of non-victim cases differs by less than 2.5\%, indicating that malicious \mlcs generalize comparably well as their benign counterparts.

\item {\em Easiness.} We show that to launch such attacks, the adversary only needs access to a small fraction (e.g., less than 22\% in the case of word-embedding \mlcs) of the training data used by the \ml system developers, which can often be obtained from public domains, meanwhile requiring no prior knowledge about how the host \ml systems are built or trained.
\end{myitemize}
%

We also provide analytical justification for the success of \mlc-based attacks, which points to the fundamental characteristics of today's \ml models: high dimensionality, non-linearity, and non-convexity, allowing the adversary to precisely alter an \ml system's behavior on a singular input without significantly affecting other inputs. This analysis leads to the conclusion that the security risks of third-party \mlcs are likely to be fundamental to \ml systems.

We further discuss potential countermeasures against \mlc-based attacks. Although it is easy to conceive high-level mitigation strategies such as more principled practice of \mlc integration, it is challenging to concretely implement such policies in specific \ml systems. For example, vetting the integrity of an \mlc for potential logic bombs amounts to searching for abnormal alterations induced by this \mlc in the feature space, which presents non-trivial challenges due to the feature space dimensionality and the \mlc model complexity. Therefore, effective defense against \mlc-based attacks is a future research topic with strong practical relevance.

\subsubsection*{\bf Contributions} In summary, we conduct the first in-depth study on the practice of reusing third-party \mlcs in building and operating \ml systems and reveal its profound security implications. Our contributions are summarized as follows.

\begin{myitemize}
\item We empirically study the status quo of reusing third-party \mlcs in developing \ml systems and show that a large number of \ml systems are built upon popular \mlcs. This study suggests that these \mlcs, once adversarially manipulated, entail immense threats to the security of a range of \ml systems.

\item We present a broad class of logic-bomb attacks and implement them on word-embedding and neural-network \mlcs. By empirically evaluating them on two state-of-the-art \ml systems in the healthcare domain, we show the effectiveness, evasiveness, and easiness of such attacks.

\item We analyze the root cause of the success of \mlc-based attacks and discuss possible mitigation strategies and their technical challenges. This analysis suggests the necessity of a significant improvement of the current practice for \mlc integration and use in developing \ml systems.
\end{myitemize}

\subsubsection*{\bf Roadmap} The remainder of the paper proceeds as follows. \myref{sec:back} studies the empirical use of \mlcs in
\ml system development; \myref{sec:attack} gives an overview of logic-bomb attacks; \myref{sec:attack1} and \myref{sec:attack2} conduct case studies of such attacks against real \ml systems; \myref{sec:discussion} provides analytical justification for the success of \mlc-based attacks and discusses potential countermeasures; \myref{sec:literature} surveys relevant literature; \myref{sec:conclusion} concludes the paper and points to future directions.



%

\section{Background}
\label{sec:back}

We begin with an empirical study on the current status of \mlcs used in building \ml systems. For ease of exposition, we first introduce a set of fundamental concepts used throughout the paper.

\subsection{Modular Development of ML Systems}

While our discussion is generalizable to other settings, in the following, we focus primarily on classification tasks in which an \ml system categorizes a given input into one of predefined classes. For instance, a skin cancer screening system takes images of skin lesions as inputs and classify them as benign lesions or malignant skin cancers~\cite{Esteva:2017:nature}.

An end-to-end classification system often comprises a number of modules, each implementing a modular functionality (e.g., feature selection, classification, and visualization). To simplify our discussion, we focus on two core modules, {\em feature extractor} and {\em classifier}, which are found across most \ml systems.

\begin{figure}
\centering
\epsfig{file=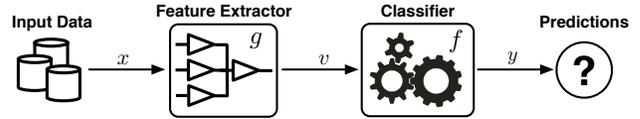, width=85mm}
\caption{Simplified workflow of a typical \ml system (only the inference process is shown). \label{fig:flow}}
\end{figure}

A feature extractor models a function $g: \mathcal{X} \rightarrow \mathcal{V}$, which projects an {\em input vector} $\vx \in \mathcal{X}$ to a {\em feature vector} $\vv{v} \in \mathcal{V}$. In general, $g$ is a many-to-one mapping. For example, $\vx$ may be an English sentence, from which $g$ extracts the counts of individual words (i.e., ``bag-of-words'' features). Meanwhile, a classifier models a function $f: \mathcal{V} \rightarrow  \mathcal{Y}$, which maps a given feature vector $\vv{v}$ to a nominal variable ranging over a set of classes $\mathcal{Y}$. The entire \ml system is thus a composition function $f \circ g: \mathcal{X} \rightarrow \mathcal{Y}$, which is illustrated in Figure\,\ref{fig:flow}. In this paper, we primarily focus on \mlcs that implement feature extractors, due to their prevalent use.

We consider \ml systems obtained via supervised learning. The training algorithm takes as input a training set $\mathcal{T}$, of which each instance $(\vx, y)$ comprises an input and its ground-truth class label. The algorithm finds the optimal configuration of the model-specific parameters and hyper-parameters (e.g., the kernel-type of an SVM classifier) by optimizing an objective function $\ell(f\circ g (\vx), y)$ for $(\vx, y) \in \mathcal{T}$ (e.g., the cross entropy between the ground-truth class labels and the outputs of $f\circ g$).

 The system developer may opt to perform {\em full-system tuning} to train both the feature extractor and the classifier. In practice, because the process of feature extraction is often non-specific to concrete data or tasks (e.g., extracting features from images of natural objects is similar to that from images of artifacts), feature extractors that are pre-trained on sufficiently representative datasets (e.g., {\sf ImageNet}~\cite{ILSVRC15}) are often reusable in a range of other domains and applications~\cite{Yosinski:2014:nips}. Therefore, the system developer may also choose to perform {\em partial-system tuning} to only train the classifier, while keeping the feature extractor intact.


\subsection{Status Quo of MLCs}

To understand the usage of \mlcs in \ml systems, we conduct an empirical study on {\sf GitHub}~\cite{github}, the world's largest open-source software development platform. We examine a collection of projects, which had been active (i.e., {\sf committed} at least 10 times) in 2016.

Among this collection of projects, we identify the set of \ml systems as those built upon certain \ml techniques. To do so, we analyze their {\sf README.md} files and search for \ml-relevant keywords, for which we adopt the glossary of~\cite{glossary}. This filtering results in 27,123 projects. To validate the accuracy of our approach, we manually examine 100 positive and 100 negative cases selected at random, and find neither false positive nor false negative cases.

\begin{table}{\small
  \centering
\begin{tabular}{lc}
{\bf MLC}             &           {\bf \# Projects} \\
\hline
\hline
\word~\cite{2013:Mikolov:arxiv}                   &              928\\
\glove~\cite{Pennington:2014:emnlp}                 &              577\\
\hline

\gnet~\cite{Szegedy:2015:cvpr}                &             466\\
 \anet~\cite{Krizhevsky:2012:nips}             &               303\\
\icep~\cite{Szegedy:2015:arxiv}                     &          190\\
\rnet~\cite{He:2015:arxiv}                 &        341\\
\vgg~\cite{Simonyan:2014:arxiv}   &  931\\
\hline
\end{tabular}
\caption{Usage of popular \mlcs in ative {\sf GitHub} projects. \label{tab:use}}}
\end{table}

%

To be succinct, we select a set of representative \mlcs and investigate their usage in the collection of \ml systems. We focus on two types of \mlcs: (i) word-embedding \mlcs (e.g., \word~\cite{2013:Mikolov:arxiv} and \glove~\cite{Pennington:2014:emnlp}), which transform words or phrases from natural language vocabularies to vectors of real numbers, and (ii) neural-network \mlcs (e.g., \gnet~\cite{Szegedy:2015:cvpr}, \anet~\cite{Krizhevsky:2012:nips}, \icep~\cite{Szegedy:2015:arxiv}, \rnet~\cite{He:2015:arxiv}, and \vgg~\cite{Simonyan:2014:arxiv}) learn high-level abstractions from complex data. Pre-trained word-embedding \mlcs are often used to reconstruct linguistic contexts of words in natural language processing (NLP), while one of neural-network \mlcs' dominant uses is to extract features from images data. Table~\ref{tab:use} summarizes the usage statistics of these \mlcs. It is observed that 3,738 projects use at least one of these \mlcs, accounting for 13.8\% of all the active \ml projects.

\begin{figure}
\centering
\epsfig{file=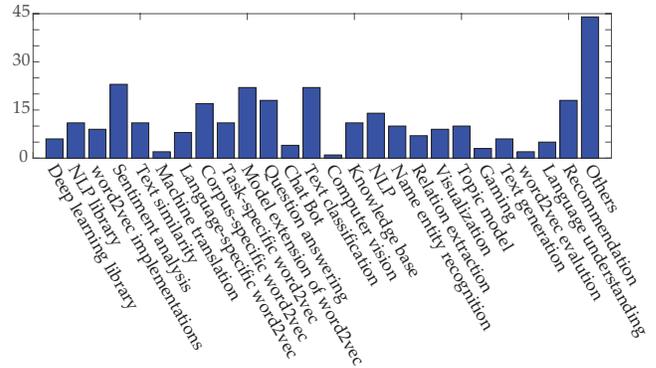, width=85mm}
\caption{Number of \word-based {\sf  GitHub} projects in different application categories. \label{fig:category}}
\end{figure}

We further investigate the application domains of a particular \mlc, \word. The results are summarized in Figure\,\ref{fig:category}. It is shown that \word is employed as a basic building block in a variety of projects ranging from general-purpose \ml libraries to domain-specific applications (e.g., Chatbot). Similar phenomena are observed with respect to other popular \mlcs. It is therefore conceivable that, given their widespread use, popular \mlcs, once adversarially manipulated, entail immense threats to the security of a range of \ml systems.

\subsection{Attack Vectors}

We consider two major channels through which potentially harmful \mlcs may penetrate and infect \ml systems.

First, they may be incorporated during \ml system development. Due to the lack of standardization, a number of variants of the same \mlc may exist in the market. For example, besides its general-purpose implementations, \word has a number of domain-specific versions (e.g., {\sf BioVec}~\cite{Asgari:2015:plos}). Even worse, potentially harmful \mlcs may be nested in other \mlcs. For example, an ensemble feature extractor may contain multiple ``atomic'' feature extractor \mlcs.
Under the pressure of releasing new systems, the developers often lack sufficient time or effective tools to vet malicious \mlcs.

Second, they may also be incorporate during \ml system maintenance. Given their dependency on training data, \mlcs are subject to frequent updates as new data becomes available. For example, the variants of \glove include .6B, .27B, .42B, and .840B~\cite{Pennington:2014:emnlp}, each trained using an increasingly larger dataset. As {\em in vivo} tuning of an \ml system typically requires re-training the entire system, the system developers are tempted to simply incorporate \mlc updates without in-depth inspection.

Both scenarios above can be modeled as follows. The adversary crafts a harmful feature extractor \mlc $\hat{g}$ (e.g., by slightly modifying a genuine \mlc $g$). The system developer accidentally obtains and incorporates $\hat{g}$, in conjunction of a classifier $f$, to build a  classification system; after the integration, the system developer may opt to train the entire system $f\circ \hat{g}$ (i.e., full-system tuning) or train the classifier $f$ only (i.e., partial-system tuning).

\section{Logic Bomb Attack}
\label{sec:attack}

Harmful \mlcs, once integrated into \ml systems, are able to significantly deviate the host systems from their expected behaviors. Next we present a broad class of logic-bomb attacks, in which malicious \mlcs force their host systems to malfunction in a predictable manner once certain predefined conditions are met (e.g., misclassification of particular inputs).
%

\subsection{Adversary Model}

Without loss of generality, we assume that the adversary attempts to trigger the \ml system to misclassify a particular input $\vxs$ into a desired target class $\vys$. For example, $\vxs$ can be the facial information of an unauthorized personnel, while $\vys$ is the decision of granting access. We refer to the instance $(\vxs, \vys)$ as a logic bomb, which is triggered when $\vxs$ is fed as input to the \ml system.


Recall that an \ml system essentially models a composition function $f\circ g: \mathcal{X} \rightarrow \mathcal{Y}$.
To achieve her goal, the adversary crafts a logic bomb-embedded \mlc $\hat{g}$ such that $f\circ \hat{g}$ classifies $\vxs$ as $\vys$ with high probability. Meanwhile, to evade possible detection mechanisms, the adversary strives to maximize the logic bomb's evasiveness by making $\hat{g}$ highly indistinguishable from its genuine counterpart $g$. More specifically,
\begin{myitemize}
\item Syntactic indiscernibility - $g$ should resemble $\hat{g}$ in terms of syntactic representation. Letting $g$ ($\hat{g}$) denote both a \mlc and its model parameters (i.e., encoded as a vector), then $g\approx \hat{g}$.
\item Semantic indiscernibility -  $g$ and $\hat{g}$ should behave similarly in terms of classifying inputs other than $\vxs$, that is,
$f\circ g(\vx) \approx f\circ \hat{g}(\vx)$ for $\vx \neq \vxs$.
\end{myitemize}


To make the attacks practical, we make the following assumptions. The adversary has no prior knowledge about the building (e.g., the classifier $f$) or the training (e.g., full- or partial-system tuning) of the host \ml system, but has access to a reference set $\mathcal{R}$, which is a subset of the training set $\mathcal{T}$ owned by the system developer. We remark that this assumption is realistic, for the adversary can often obtain some relevant training data from public domains, while the system developer may have access to certain private data inaccessible by the adversary.

\begin{algorithm}{\small
\KwIn{logic bomb $(\vxs, \vys)$, reference set $\mathcal{R}$, genuine \mlc $g$}
\KwOut{bomb-embedded \mlc $\hat{g}$}

\tcp{$\Delta$: perturbation operations}
$\hat{g} \leftarrow g$, $\Delta \leftarrow \emptyset$\;

\While{$\Delta$ satisfies indiscernibility conditions w.r.t. $\mathcal{R}$}{
apply $\Delta$ to $\hat{g}$\;
$\Delta \leftarrow$ perturbation to $\hat{g}$ to maximize $\mu(\vxs, \vys)$\;
}
return $\hat{g}$\;
\caption{Crafting logic bomb-embedded \mlc \label{alg:bomb}}}
\end{algorithm}

\subsection{A Nutshell View}
\label{sec:nutshell}

For ease of exposition, below we play the role of the adversary to describe the attack model. At a high level, we craft $\hat{g}$ by carefully perturbing its benign counterpart $g$. For instance, in the case of a word-embedding \mlc, we modify the embedding mappings of a subset of the words, while in the case of a neural-network \mlc, we modify a subset of the model parameters of the neural network, but without changing its architecture.

To guide the perturbation, we rely on a {\em membership} function $\mu(\vx, y)$ that measures the proximity of an input $\vx$ to a class $y$, with reference to the current $\hat{g}$. The ideal form of $\mu(\vx, y)$ is obviously $\ell(f\circ \hat{g}(\vx), y)$ with $\ell(\cdot)$ being the objective function used by the system developer to train the system. However, without knowledge about the building or the training of the host system, we need to approximate this membership function. We consider the following two strategies.

\subsubsection*{\bf Unsupervised Strategy} We may leverage the similarity of $\vxs$ and the inputs in the reference set $\mathcal{R}$ to approximate $\mu(\cdot, \cdot)$. Let $\mathcal{R}$ be divided into different subsets $\{\mathcal{R}_y\}$, each comprising the inputs in one class $y$. Let $\vv{x_y}$ denote the centroid of $y$:
\begin{equation}
  \label{eq:center}
  \vv{x_y} = \frac{1}{|\mathcal{R}_y|}\sum_{\vx \in \mathcal{R}_y} \vx
\end{equation}
One possible form of the membership function can be defined as:
\begin{equation}
   \label{eq:member}
\mu(\vx, y) = \frac{\exp(-|| \hat{g}(\vx) - \hat{g}(\vv{x_y}) ||)}{\sum_{y'\in \mathcal{Y}} \exp(-|| \hat{g}(\vx) - \hat{g}(\vv{x_{y'}}) ||)}
\end{equation}
where $||\cdot||$ is a properly chosen norm. Note that this function is normalized: $\sum_{y\in \mathcal{Y}} \mu(\vx, y) = 1$.

\subsubsection*{\bf Supervised Strategy}

Note that the membership function defined in Eq\,(\ref{eq:member}) implicitly treats each feature with equal importance. This simplification may not work well for high-dimensional feature spaces (e.g., tens of thousands of dimensions). Instead, we may resort to a {\em surrogate} classifier $\hat{f}$ to approximate $f$ and use the objective function
$\ell(\hat{f}\circ \hat{g}(\vx), y)$ as the membership function.

A variety of off-the-shelf classifiers may serve this purpose (e.g., logistic regression, support vector machine, multilayer perceptron). In our empirical study (\myref{sec:attack2}), we find that the concrete form of $\hat{f}$ is often immaterial, as long as it provides informative classification results, even with lower
accuracy than $f$.



\subsubsection*{\bf Attack Model}

Algorithm~\ref{alg:bomb} sketches our attack model.
We (iteratively) perturbs $\hat{g}$ to maximize the membership of $\vxs$ with respect to the target class $\vys$, $\mu(\vxs, \vys)$. At each step, we also ensure that the syntactic and semantic indiscernibility conditions are met. While the high-level idea seems simple, implementing such attacks on concrete \mlcs presents non-trivial challenges, including (i) how to perturb the \mlc model to maximize $\mu(\vxs, \vys)$, (ii) how to guarantee the indiscernibility constraints, and (iii) how to strike the balance between these two criteria.

Next we show the concrete implementation of Algorithm~\ref{alg:bomb} in the cases of word-embedding \mlcs (\myref{sec:attack1}) and neural-network \mlcs (\myref{sec:attack2}) and empirically validate the feasibility of logic-bomb attacks against real \ml systems.


\section{Attacking Word-Embedding MLC}
\label{sec:attack1}


\subsection{Word-Embedding MLC}

Word embeddings represent a family of feature extractors, which project words in natural languages to vectors of real numbers. Conceptually it involves a mathematical embedding from a space with one dimension per word to a continuous vector space with much lower dimension. For example, in a popular \word implementation by T. Mikolov {\em et al.}~\cite{2013:Mikolov:arxiv}, each of 3 million words is mapped to a 300-dimensional vector.

\begin{figure}
\centering
\epsfig{file=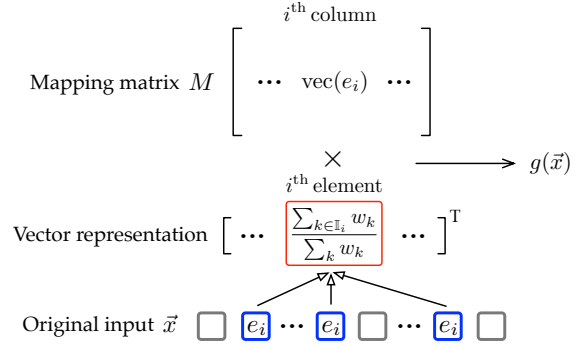, width=75mm}
\caption{Anatomy of word-embedding feature extractor. \label{fig:aggre}}
\end{figure}

One may imagine a pre-trained word embedding as a dictionary, mapping each word $e_i$ in a vocabulary $\mathcal{E} = \{e_i\}$ to its embedding vector $\mathsf{vec}(e_i)$ by a function $\mathsf{vec}(\cdot)$. This function can be generated using
neural networks~\cite{2013:Mikolov:arxiv}, dimensionality reduction~\cite{Levy:2014:nips}, or probabilistic models~\cite{Globerson:2007:jmlr}. We represent a word embedding by a matrix $M$ wherein the $i^{\textrm{th}}$ column encodes $\mathsf{vec}(e_i)$, as shown in Figure\,\ref{fig:aggre}.

Each input $\vx$ comprises an ordered sequence of words: $(x_1, x_2, \ldots)$, with $x_\sk \in \mathcal{E}$ as the $k^\textrm{th}$ word of $\vx$ and $k$ as its {\em index}. The feature extractor $g(\cdot)$ is often defined as computing the weighted average of the embedding vectors of all the words:
\begin{equation}
	\label{eq:word}
g(\vx) = \frac{1}{\sum_\sk w_\sk}\sum_\sk w_\sk \cdot \mathsf{vec}(x_\sk)
\end{equation}
where $w_\sk$ is the weight of the $k^\textrm{th}$ word in the sequence. For instance, with $w_\sk = \gamma^{1-\sk}$ for $0 <  \gamma \leq 1$, $\vx$'s feature vector is defined as the ``time-decaying'' average of all the embedding vectors.

As illustrated in Figure\,\ref{fig:aggre}, to simplify Eq.(\ref{eq:word}), we may vectorize $\vx$ as an $|\mathcal{E}|$-dimensional vector: the $\order{i}$ element is defined as $\sum_{\sk \in \mathbb{I}_i} w_\sk/\sum_{\sk} w_\sk$, with $\mathbb{I}_i$ representing the set of indices where the word $e_i$ appears in $\vx$. Intuitively, this transformation aggregates the weight of each distinct word in $\vx$. With a little abuse of notation, below we use $\vx$ to refer to both an input and its vector representation. Under this setting, the computation of $\vx$'s feature vector is simplified as matrix-vector multiplication:
\begin{equation}
  \label{eq:mapping}
g(\vx) = M \cdot \vx
\end{equation}
 which is then fed as input to the classifier. 

\subsection{Attack Implementation}
As indicated in Eq.(\ref{eq:mapping}), the core of word embedding is the mapping matrix $M$.
We thus attempt to embed the logic bomb $(\vxs, \vys)$ by carefully perturbing $M$ while minimizing the impact on inputs other than $\vxs$. This attack can be modeled as adding a perturbation matrix $E$ to $M$: $\hat{M}  = M + E$.

Without loss of generality, we make the following assumptions: (i) the \ml system performs binary classification, i.e., $\mathcal{Y} = \{-, +\}$; (ii) the adversary intends to trigger the system to misclassify $\vxs$ as `+' (from `-'); (iv) in the reference set $\mathcal{R}$, $\vv{x_\splus}$ and $\vv{x_\sminus}$ respectively represent the centroid of the class `+' and `-' ({\em cf.} Eq.(\ref{eq:center})).

\subsubsection*{\bf Formulation}

We adopt the unsupervised strategy (\myref{sec:nutshell}) to approximate the membership function $\mu(\cdot, \cdot)$ and translate the perturbation process into an optimization framework. Specifically,
\begin{myitemize}
\item First, maximizing the membership of $\vxs$ with respect to the class `+' ({\em cf.} Eq.(\ref{eq:member})) is equivalent to minimizing the quantity of
$|| \hat{M}\cdot ( \vxs - \vv{x_\splus} )   ||   -|| \hat{M}\cdot ( \vxs - \vv{x_\sminus} )   ||$.
\item Second, as $M$ and $\hat{M}$ differ by the perturbation matrix $E$, we enforce the syntactic indiscernibility by bounding its norm $||E||$, which is equivalent to adding $||E||$ as a regularization term to the objective function, with a parameter $\lambda$ indicating the importance of this constraint.

\item Third, to enforce the semantic indiscernibility, besides bounding $||E||$, we also ensure that for each input $\vx \in \mathcal{R}$, its feature vector varies only slightly before and after the perturbation, i.e., $  || g(\vx) - \hat{g}(\vx) || = ||E \cdot \vx ||\leq \delta$ ($\delta$ as a threshold).
\end{myitemize}

Putting everything together, the perturbation operation in Algorithm~\ref{alg:bomb} amounts to
solving the following problem:
\begin{displaymath}
\begin{array}{cc}
\mathsf{minimize}_{E} & || \hat{M}\cdot ( \vxs - \vv{x_\splus} )   ||   -|| \hat{M}\cdot ( \vxs - \vv{x_\sminus} )   ||  + \lambda \cdot ||E||\\[2pt]
\mathsf{s.t.} &
||E \cdot \vx ||\leq \delta \; \mathsf{for} \; \vx \in \mathcal{R}
\end{array}
\end{displaymath}

Note that it is straightforward to generalize this formulation to the case of multiple target inputs (details in Appendix A).

\subsubsection*{\bf Optimization}

As different norms are within a constant factor of one another, without loss of generality, we adopt $l_2$ norm (i.e., Frobenius norm for matrices) and consider exploring other norms as our ongoing research. The optimization problem above is instantiated as a {\em quadratically constrained quadratic program} (QCQP), whose convexity depends on the concrete data values. Non-convex QCQPs are NP-hard in general. We thus resort to numerical solutions and extend the solver in~\cite{Park:2017:arxiv} to solve this problem.

For the computational efficiency and the attack's evasiveness, it may be also desirable to constrain the number of embedding vectors in $M$ to be perturbed (i.e., the number of nonzero vectors in $E$). In our current implementation, we select the subset of embedding vectors to perturb using a heuristic rule: we pick those vectors with lowest norm, which tend to correspond to ``unimportant'' words.

%


\subsection{A Case-Study System}

To validate the feasibility of our attack model, we perform an empirical study on a real \ml system in the healthcare domain, ``Sequential Phenotype Predictor'' ({\sf SPP})~\cite{farhan:2016:jmir}.

\begin{figure}[h]
	\centering
	\epsfig{file=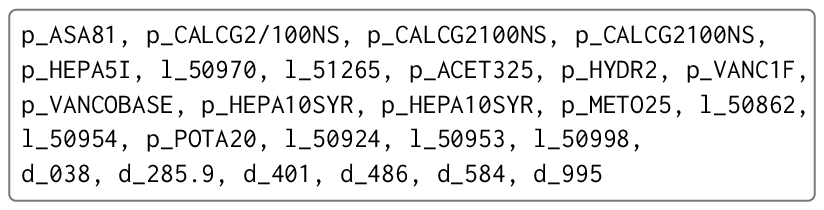, width=80mm}
\caption{Example of a patient's EHR \label{fig:ehr}}
\end{figure}

At a high level, {\sf SPP} is a disease early prognosis system, which predicts the likely course of a given patient's disease, in particular, the diagnosis of her next admission, based on her diagnosis and treatment history in previous admissions. The system considers a patient's electronic health record (EHR) as a sequence of medical events (e.g., lab tests, prescriptions, diagnosis), each analogous to a word in a sentence. An example is given in Figure\,\ref{fig:ehr}: all the medical event are encoded using the ICD9 code\footnote{ICD9: The 9th Revision of the International Classification of Diseases (www.who.int)}, with lab tests prefixed with {\em ``l''}, prescriptions with {\em ``p''} and diagnosis with {\em ``d''}.


The feature extractor $g$ of {\sf SPP} is a domain-specific {\word} \mlc that is pre-trained on a large EHR corpus (e.g., OpenMRS\footnote{OpenMRS: http://openmrs.org}). Specifically, the vocabulary $\mathcal{E}$ consists of 2,761 medical events and $g$ projects a given patient's medical event sequence $\vx$ to a 100-dimensional feature vector $g(\vx)$ according to Eq.(\ref{eq:word}). For each possible diagnosis (e.g. chronic kidney failure), {\sf SPP} trains a binary classifier $f$ (e.g., logistic regression), which, integrated with $g$, forms a classification system $f \circ g$ to predict the probability that a given patient will be diagnosed with this disease.
\begin{table}[h]{\small
  \centering
\begin{tabular}{|c|c|c|}
	\hline
{\bf ICD9}     &  {\bf Disease}    &       {\bf Positive Rate (\%)} \\
\hline
\hline
d\_427 & Cardiac dysrhythmias &  37.0\\
d\_428 & Heart failure & 35.7 \\
d\_250 & Diabetes mellitus & 32.0\\
d\_272 & Disorders of lipoid metabolism & 25.6\\
\hline
\end{tabular}
\caption{List of diseases in our case study. \label{tab:disease}}}
\end{table}
\begin{figure*}
	\epsfig{file=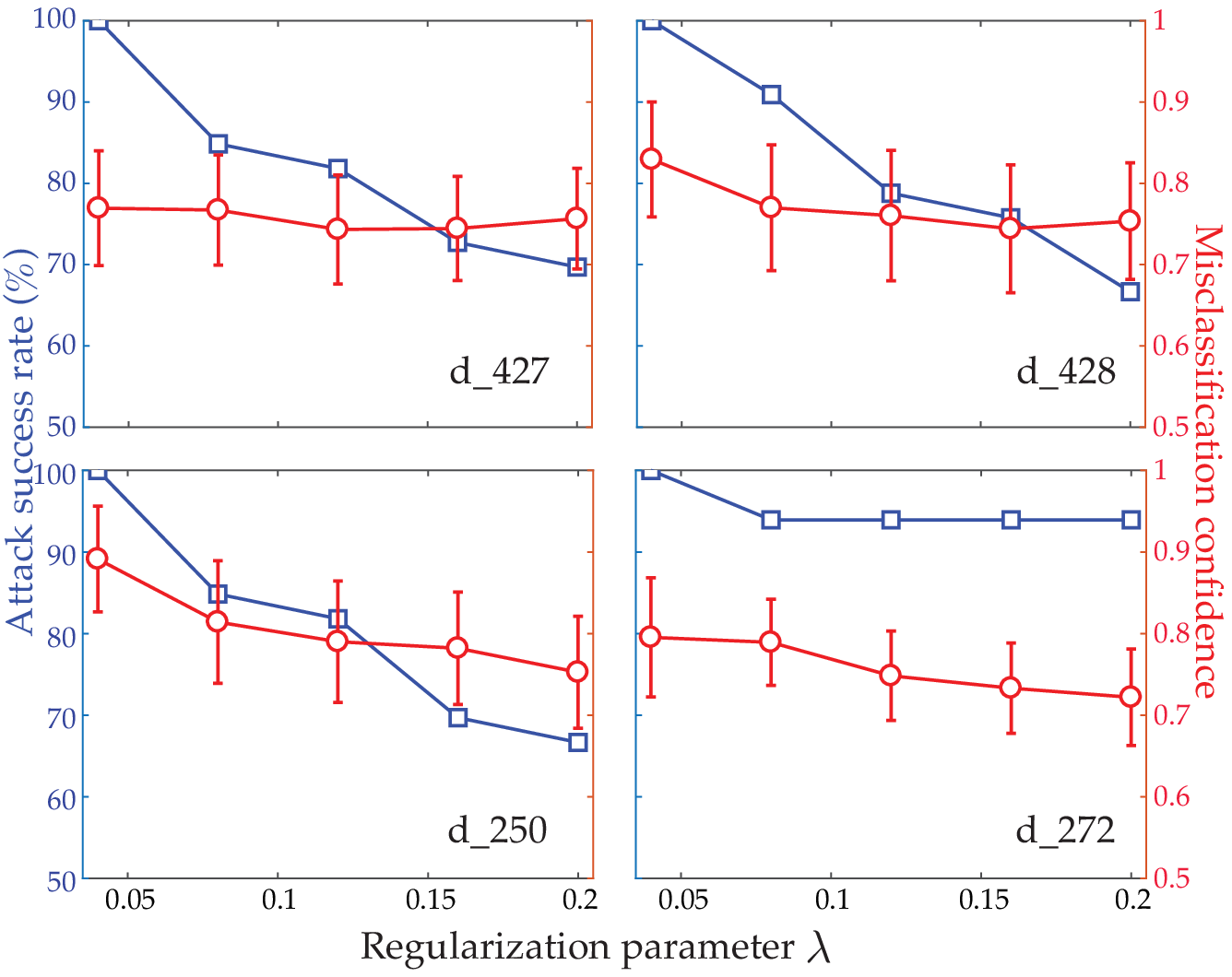, width=88mm}
	\epsfig{file=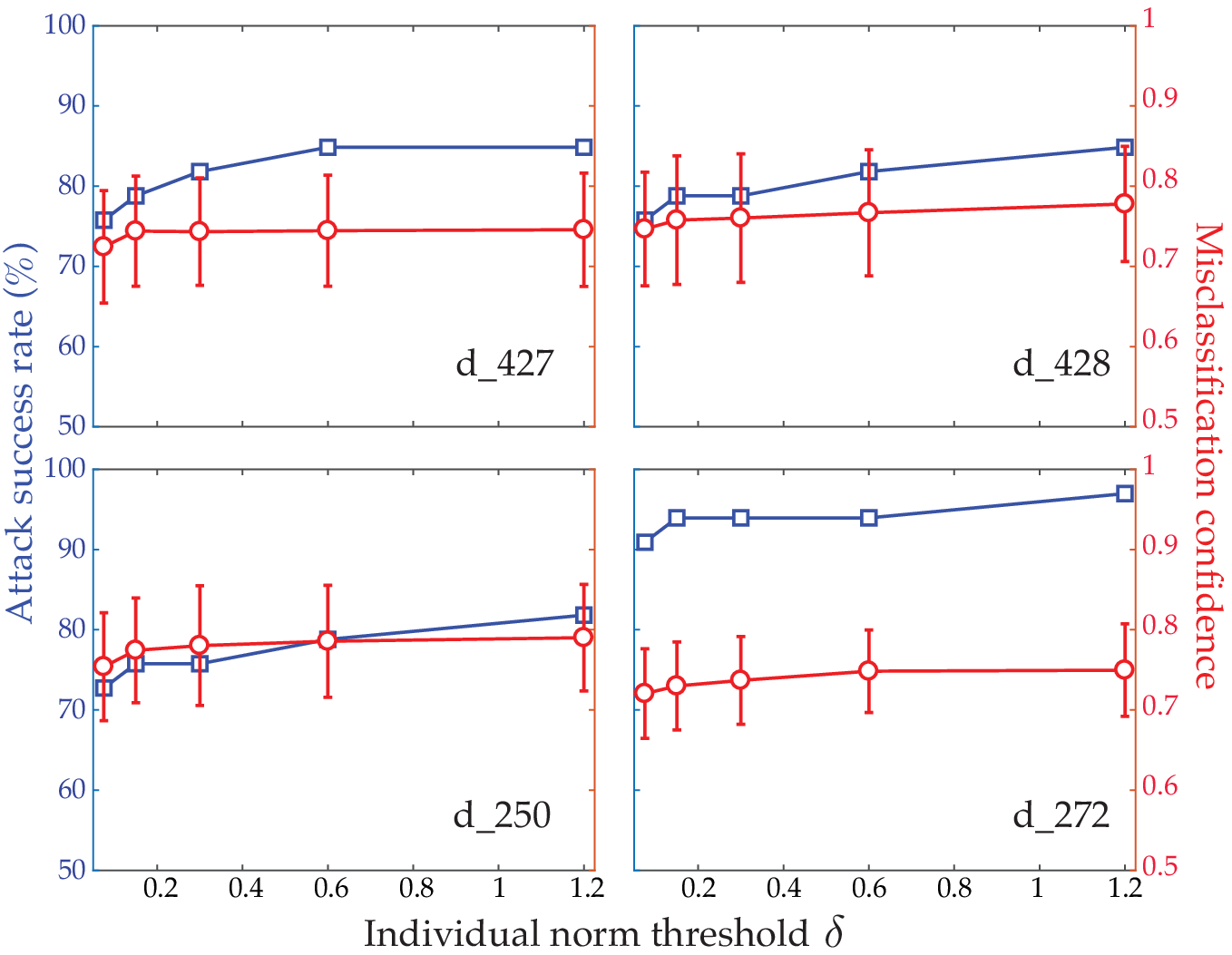, width=88mm}
\caption{Effectiveness of logic-bomb attacks versus indiscernibility parameters ($\lambda$ and $\delta$). \label{fig:successrate}}
\end{figure*}

\subsubsection*{\bf Dataset}

We use the MIMIC-III dataset~\cite{mimic}, which constitutes the EHRs of 5,642 patients, each with 192.7 medical events on average. We use 80\% of the dataset as the training set $\mathcal{T}$ and the remaining 20\% as the validation set $\mathcal{V}$. We assume: (i) the system developer uses the entire $\mathcal{T}$ to train the system; (ii) the adversary has access to a subset of $\mathcal{T}$ as the reference set $\mathcal{R}$, but without knowledge about the classifier in the system; (iii) she intends to force the misclassification of a target patient $\vxs$ in $\mathcal{V}$, but without information about other patients in $\mathcal{V}$.

In the study, we consider the diagnosis of four severe diseases, with their information summarized in Table\,\ref{tab:disease}. For each disease, {\sf SPP} trains a binary classifier (e.g., logistic regression).

%
%

\subsection{Empirical Evaluation}

\label{sec:attack1exp}

\subsubsection*{\bf Setting}

For comparison, we first build a baseline system $f\circ g$ that integrates a genuine feature extractor $g$ and a classifier $f$. In each trial of the attack, we randomly select one input $\vxs \in \mathcal{V}$, which is correctly classified by the baseline system, and craft a malicious \mlc $\hat{g}$ to embed the logic bomb $(\vxs, \vys)$ ($\vys$ is the misclassification). We then compare the performance of the infected system $f \circ \hat{g}$ against the baseline system $f\circ g$.

For each set of experiments, we run 50 trials. The default classifier is a logistic regression model. The default setting of key parameters is as follows:  the regularization parameter $\lambda = 0.12$, the vector norm threshold $\delta = 0.3$, the number of perturbed vectors $n = 20$, and the size of reference set $|\mathcal{R}| = 1000$.

\subsubsection*{\bf Summary} We design our study to evaluate logic-bomb attacks from three aspects:

\begin{myitemize}
\item Effectiveness: we show that with 100\% success rate, logic-bomb attacks succeed in forcing the host \ml system to misclassify target inputs with average confidence above 0.76.

\item Evasiveness: we show that it is possible to achieve over $75\%$ attack success rate via perturbing only 0.7\% of the embedding vectors, each parameter with distortion less than 0.004, while the impact on the classification of other inputs is below $0.6\%$.

\item Easiness: we show that the adversary only needs to access about 22\% (in some cases less than 2.2\%) of the training data to achieve  close to 80\% attack success rate, regardless of the building or training of the host \ml system.
\end{myitemize}

Next we detail our studies.

\subsubsection*{\bf Effectiveness}

In the first set of experiments, we evaluate the effectiveness of logic-bomb attacks against the case-study system. We use two metrics: (i) attack success rate, which measures the likelihood that the system is fooled to misclassify the target input:
\begin{displaymath}
\textsf{Attack success rate} = \frac{\textsf{\# misclassifications}}{\textsf{\# trials}}
\end{displaymath}
and (ii) {\em misclassification confidence}, which is the probability assigned to the misclassified target input by the system. Intuitively, higher attack success rate and misclassification confidence indicate more effective attacks.

Figure\,\ref{fig:successrate} illustrates the effectiveness of logic-bomb attacks as a function of the indiscernibility parameters, where
the regularization parameter $\lambda$ varies from 0.04 to 0.2, while the individual norm threshold $\delta$ grows from 0.075 to 1.2 (in geometric progression). In both cases, we consider the systems built for the diagnosis of four diseases listed in Table~\ref{tab:disease}.

%
%

We have the following observations. First, logic-bomb attacks are highly effective against the case-study system. For example, under $\lambda = 0.04$, the attacks achieve 100\% success rate with average misclassification confidence above 0.76 across all the settings. Second, there exists a fundamental tradeoff between the attack's effectiveness and its stealthiness. Intuitively, a larger perturbation amplitude implies more manipulation room for the adversary. For example, as $\delta$ increases from 0.075 to 1.2, the attack success rate increases by about 10\% in the diagnosis system for d\_427. However, even under fairly strict indiscernibility constraints, logic-bomb attacks remain effective. For example, under $\lambda = 0.2$ (which corresponds to bounding the distortion to each parameter by 0.004, as will be revealed shortly), the adversary is still able to force the system to misclassify more than 70\% of the target inputs in all the cases. Third, compared with the constraint on the perturbation matrix ($\lambda$), the constraints on the individual inputs ($\delta$) seem have less influence on the attack's effectiveness. This may be explained as follows: $\lambda$ directly bounds the perturbation matrix $E$, while $\sigma$ bounds $E$ through the matrix-vector multiplication $||E \cdot \vx||$; due to the sparseness of $\vx$ (i.e., each patient is associated with a limited number of medical events), the bound on $||E \cdot \vx||$ is much less effective.

\begin{figure}
	\epsfig{file=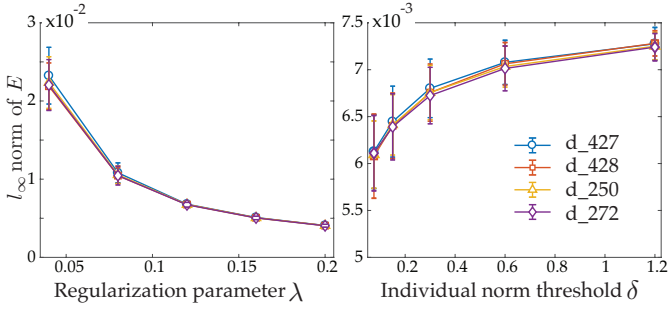, width=88mm}
\caption{Syntactic evasiveness of logic-bomb attacks versus indiscernibility parameters ($\lambda$ and $\delta$). \label{fig:flipa}}
\end{figure}

\begin{table*}{\small

\begin{tabular}{|c|c|c|c|c|c|c|c|c|c|}

\hline
\multirow{2}{*}{\bf Classifier} & \multirow{2}{*}{\bf Disease} &  \multicolumn{4}{c}{\bf Attack Success Rate (\%)} & \multicolumn{4}{|c|}{\bf Classification Flipping Rate (\%)}\\
\cline{3-10}
& & $|\mathcal{R}| = 100$ & $|\mathcal{R}| = 200$ & $|\mathcal{R}| = 500$ & $|\mathcal{R}| = 1000$  & $|\mathcal{R}| = 100$ & $|\mathcal{R}| = 200$ & $|\mathcal{R}| = 500$ & $|\mathcal{R}| = 1000$ \\
\hline
\hline
\multirow{4}{*}{\bf LR} & {\em d\_427} & 75.8 & 78.8 & 81.8 & 84.8 & 0.98$\pm$0.29 & 0.81$\pm$0.17 & 0.76$\pm$0.17 & 0.69$\pm$0.10 \\
 & {\em d\_428} & 72.7 & 72.7 & 78.8 & 84.8 & 1.03$\pm$0.29 & 0.90$\pm$0.24 & 0.86$\pm$0.15 & 0.67$\pm$0.14 \\
 & {\em d\_250} & 69.7 & 75.8 & 81.8 & 84.8 & 1.60$\pm$0.32 & 1.26$\pm$0.24 & 0.91$\pm$0.19 & 0.84$\pm$0.14 \\
 & {\em d\_272} & 87.9 & 90.9 & 90.9 & 93.9 & 1.16$\pm$0.40 & 1.08$\pm$0.32 & 0.96$\pm$0.35 & 0.89$\pm$0.20 \\
\hline
\hline
\multirow{4}{*}{\bf SVM} & {\em d\_427} & 75.8 & 78.8 & 81.8 & 81.8 & 1.13$\pm$0.27 & 0.80$\pm$0.16 & 0.77$\pm$0.19 & 0.66$\pm$0.09 \\
 & {\em d\_428} & 72.7 & 72.7 & 81.8 & 81.8 & 0.96$\pm$0.27 & 0.94$\pm$0.19 & 0.68$\pm$0.15 & 0.64$\pm$0.11 \\
 & {\em d\_250} & 66.7 & 69.7 & 78.8 & 81.8 & 1.45$\pm$0.28 & 1.17$\pm$0.25 & 0.93$\pm$0.23 & 0.79$\pm$0.14 \\
 & {\em d\_272} & 90.9 & 93.9 & 93.9 & 93.9 & 0.94$\pm$0.39 & 0.93$\pm$0.39 & 0.69$\pm$0.29 & 0.56$\pm$0.16 \\
\hline
\hline
\multirow{4}{*}{\bf MLP} & {\em d\_427} & 63.6 & 63.6 & 66.7 & 72.7 & 1.24$\pm$0.13 & 1.10$\pm$0.33 & 0.95$\pm$0.36 & 0.64$\pm$0.14 \\
 & {\em d\_428} & 54.5 & 63.6 & 66.7 & 78.8 & 1.28$\pm$0.20 & 1.21$\pm$0.26 & 1.21$\pm$0.23 & 1.13$\pm$0.13 \\
 & {\em d\_250} & 69.7 & 72.7 & 75.8 & 78.8 & 1.33$\pm$0.17 & 1.33$\pm$0.23 & 1.30$\pm$0.23 & 1.26$\pm$0.16 \\
 & {\em d\_272} & 78.8 & 78.8 & 81.8 & 84.8 & 1.34$\pm$0.19 & 0.47$\pm$0.25 & 0.45$\pm$0.27 & 0.43$\pm$0.09 \\
\hline

\end{tabular}}

\caption{Impact of the size of reference set $\mathcal{R}$ and the instantiation of classifier $f$ on the attack's effectiveness and evasiveness.\label{tab:resource}}

\end{table*}

\subsubsection*{\bf Evasiveness}

Next we evaluate the evasiveness of logic-bomb attacks with respect to possible detection mechanisms. We consider both syntactic and semantic indiscernibility.

In terms of syntactic indiscernibility, we measure the $l_\infty$ norm of the perturbation matrix $E$ (i.e., the largest absolute value among the elements of $E$), which represents the largest discrepancy one may find between the encoding of genuine and malicious \mlcs. Figure\,\ref{fig:flipa} shows how the syntactic indiscernibility varies with the parameters $\lambda$ and $\delta$. Observe that both $\lambda$ and $\delta$ effectively control the magnitude of $E$. For example, as $\lambda$ increases from 0.04 to 0.2, the average $l_\infty$ norm of $E$ drops from 0.023 to 0.004. It is possible to hide perturbation of such magnitude in the encoding variance across different system platforms (16-bit versus 32-bit)~\cite{Goodfellow:2014:arxiv}.

\begin{figure}
	\epsfig{file=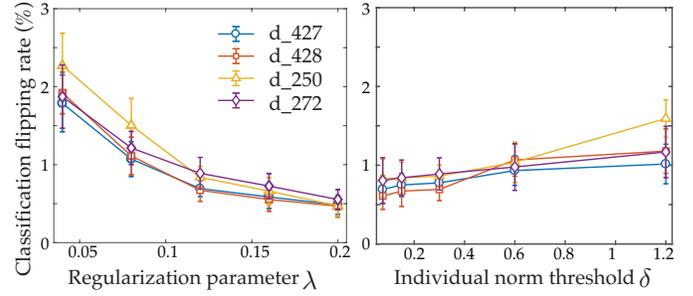, width=88mm}
\caption{Semantic evasiveness of logic-bomb attacks versus indiscernibility parameters  ($\lambda$ and $\delta$). \label{fig:flipb}}
\end{figure}

In terms of semantic indiscernibility, we measure the difference of the influence of malicious and benign \mlcs on classifying non-target inputs in the validation set $\mathcal{V}$. We introduce the metric of {\em classification flipping rate}:
\begin{displaymath}
    \textsf{Classification flipping rate} = \frac{\textsf{\# (classification}_{\hat{g}} \neq \textsf{classification}_{g})}{\textsf{\# total inputs}}
\end{displaymath}
It computes the fraction of inputs (excluding the target input $\vxs$) about which the baseline system $f\circ g$ and the infected system $f\circ \hat{g}$ disagree, reflecting the discrepancy one may find between $g$ and $\hat{g}$ in classifying non-target inputs. Figure\,\ref{fig:flipb} shows the classification flipping rate as a function of $\lambda$ or $\delta$. We have the following observations. Overall, under proper parameter settings, logic-bomb attacks incur fairly indiscernible impact (less than 1\%) on non-target inputs. Further, the flipping rate is positively correlated with the perturbation amplitude. For example, as $\delta$ increases from 0.015 to 1.2, the flipping rate gradually grows about 1\% in the case of d\_250. Finally, by adjusting $\lambda$ or $\delta$, the adversary may effectively control the classification flipping rate. For example, under $\lambda = 0.2$, the flipping rate is as low as 0.6\% across all the cases.

\subsubsection*{\bf Easiness}

To design countermeasures against logic-bomb attacks, it is essential to understand the resources necessary for the adversary to launch such attacks. Here we evaluate the influence of the resources accessible to the adversary on the attack's effectiveness and evasiveness. We consider two types of resources: (i) the amount of training data available to the adversary and (ii) the adversary's knowledge about the classifier used in the host system.

Recall that the adversary is able to access a reference set $\mathcal{R}$, a subset of the training set $\mathcal{T}$. To assess the impact of the available training data, we vary the size of $\mathcal{R}$ from 100 to 1,000. Meanwhile, to assess the impact of the instantiation of the classifier, we consider three types of classifiers, logistic regression (LR), support vector machine (SVM), and multilayer perception (MLP) with two hidden layers of size 240 and 60. We then measure the attack's effectiveness (attack success rate) and evasiveness (classification flipping rate) under varied settings. Table~\ref{tab:resource} summarizes the results. We have the following observations.

First, the amount of available training data has marginal impact on the attack's performance. For example, with $|\mathcal{R}|$ varies from 100 to 1,000, the attack success rate improves by less than 10\% across all the cases. This may be explained by the limited dependency of our attack model on $\mathcal{R}$: it is primarily used to compute the centroids of difference classes ({\em cf.} Eq.(\ref{eq:center})); provided that $\mathcal{R}$ sufficiently captures the underlying distribution of $\mathcal{T}$, its size is not critical. A similar conclusion can be drawn for the impact on the attack's evasiveness: $\mathcal{R}$ influences the classification flipping rate through the set of individual norm constraints $||E \cdot \vx||$ during the optimization process. If the inputs in $\mathcal{R}$ are representative enough, their constraints are sufficient to control the indiscernibility.


%


Second, thanks to its classifier-agnostic design, the attack model works effectively on unknown classifiers. For example, with $|R|$ = 1,000, the attack model achieves success rate close to 80\% across different classifiers. Meanwhile, observe that the attack's performance is, to a certain extent, subject to the complexity of the classifier. For example, the attack model has fairly similar performance in the cases of LR and SVM, but has over 10\% success rate drop in the case of MLP, especially under limited training data. This may be attributed to the assumption underlying the unsupervised strategy (\myref{sec:nutshell}), which treats each feature with equal importance. This approximation often performs well for low-dimensional feature spaces and varied simple classifiers (as shown in the cases of LR and SVM), but may not generalize to more complicated settings. To address this issue, in \myref{sec:attack2}, we develop a supervised attack model which is effective on some of today's most complicated classifiers (e.g., deep neural networks).

\section{Attacking Neural-Network MLC}
\label{sec:attack2}


\subsection{Neural-Network MLC} 
\begin{figure}
 \centering
 \epsfig{file=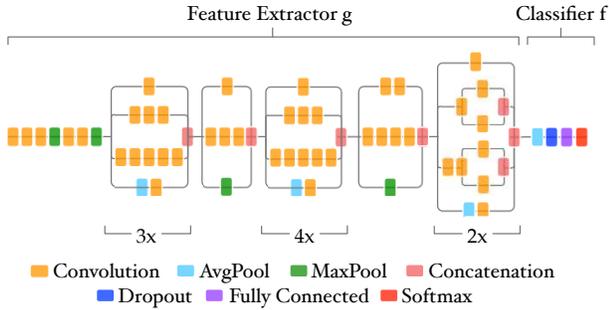, width=88mm}
 \caption{Schematic diagram of a case-study \ml system (``$n\times$'' represents a sequence of $n$ copies of the same block). \label{fig:icep}}
 \end{figure}

Deep neural networks~\cite{LeCun:2015:nature} (\dnns) represent a class of \ml algorithms to learn high-level abstractions of complex data using multiple processing layers in conjunction of non-linear transformations. One major advantage of \dnns is their ability to automatically extract intricate features from raw data without careful engineering by hand. A schematic example of \dnn is shown in Figure\,\ref{fig:icep}.

We primarily focus on the use of \dnns for image data. For a given image $\vx$, a {\dnn} \mlc, pre-trained on a sufficiently representative dataset (e.g., {\sf ImageNet}~\cite{ILSVRC15}), outputs a set of {\em feature maps}, which are collectively referred to as the feature vector $g(\vx)$.
%
 %

 \subsection{Attack Implementation}
 \label{sec:individual}

To craft a malicious \mlc $\hat{g}$ to embed a logic bomb $(\vxs, \vys)$, we selectively perturb a subset of the parameters of a benign \mlc $g$. Compared with changing the DNN architecture, modifying the \dnn parameters tends to be more detection-evasive, which can often be hidden by the encoding variance across different system platforms~\cite{Szegedy:2013:arxiv}.

While it is convenient to use the unsupervised strategy (\myref{sec:nutshell}) to approximate the membership function, we find that this method works poorly for neural-network \mlcs, largely due to the dimensionality of feature spaces (e.g., 32,768 dimensions) and the complexity of classifiers. Thus, we adopt the supervised strategy by introducing a surrogate classifier $\hat{f}$ to approximate the true classifier $f$. We find that the concrete form of $\hat{f}$ is often immaterial; for example, it can be simply a combination of a fully connected layer and a softmax layer. To initialize $\hat{f}$, we perform partial-system tuning on the reference set $\mathcal{R}$.

To ensure the syntactic indiscernibility, we enforce that the perturbation amplitude must not exceed a threshold $\epsilon$ (i.e., $||g -\hat{g}|| \leq \epsilon$). As we perturb a small number of parameters at each iteration, we use $l_\infty$ norm here.
To ensure the semantic indiscernibility, we enforce that the perturbation to $g$ should have minimal impact on the classification of non-target inputs. Apparently the key idea is to selectively
perturb a subset of parameters important for $\vxs$ but unimportant for other inputs.
However, the parameter importance in a DNN is often extremely difficult to assess due to the mutual influence and activation among interconnected neurons: a parameter may be unimportant because of the existence of some others, but it will become critical once the others are modified. Therefore, it should be more appropriate to conduct a learning process, and continuously select and modify the parameters.


 \begin{algorithm}{\small
 \KwIn{logic bomb $(\vxs, \vys)$, reference set $\mathcal{R}$, genuine \mlc $g$, thresholds $\epsilon$, $\alpha$, $\kappa$}
 \KwOut{logic bomb-embedded \mlc $\hat{g}$}

 \tcp{bootstrap: initialization of a surrogate classifier $\hat{f}$ on $\mathcal{R}$}


 $\hat{g} \leftarrow g$\;
 \tcp{candidate parameters}
 $\bar{\Theta} \leftarrow \emptyset$\;
\Do{$\bar{\Theta} \neq \emptyset$}
{

\If{$\bar{\Theta} \neq \emptyset$}{
    $k = \min(\kappa, |\bar{\Theta}|)$\;
    \tcp{top-$k$ parameters with largest positive impact}
    \For{$i = 1 $ to $k$}{
$ \theta_\sast \leftarrow \arg\max_{\theta \in \bar{\Theta}}\left|\Delta_{(\vxs, y_\sast)}^\theta \right|$\;
\tcp{MLC perturbation}
 $\theta_\sast \leftarrow \theta_\sast + \textsf{sign}\left(\Delta_{(\vxs, y_\sast)}^{\theta_\sast}\right) \cdot \epsilon$\;
    remove $\theta_\sast$ from $\bar{\Theta}$\;
    }

    %
    %
    $\bar{\Theta} \leftarrow \emptyset$\;
}

\ForEach{layer $\mathcal{L}$ of $\hat{g}$}{
$\Theta \leftarrow$ parameters of $\mathcal{L}$\;
  \tcp{negative impact threshold}
$r \leftarrow \mathsf{mean}( \{ |\theta| \}_{\theta \in \Theta}  )  - \alpha \cdot \mathsf{std}( \{ |\theta| \}_{\theta \in \Theta}  )$\;
\ForEach{$\theta \in \Theta$}{
\lIf{$|\theta| < r$}{add $\theta$ to $\bar{\Theta}$}
}

}

}

 return $\hat{g}$\;
 \caption{Crafting logic bomb-embedded \mlc \label{alg:bomb2}}}
 \end{algorithm}

\subsubsection*{\bf Positive and Negative Impact}

Let $\vs$ be the output of the surrogate classifier $\hat{f}$ (e.g., the logits of the softmax layer in Figure\,\ref{fig:icep}), wherein the element $\sigma_y$ represents the predicted probability of the given input belonging to the class $y$. We quantify the importance of a parameter $\theta$ with respect to a given input-output instance $(\vx, y)$ using a  partial derivative measure:
 \begin{equation}
   \label{eq:saliency}
 \Delta_{(\vx, y)}^\theta = \frac{\partial \sigma_{y} }{\partial \theta} - \sum_{y' \neq y} \frac{\partial \sigma_{y'} }{\partial \theta}
 \end{equation}
 where the first term quantifies the influence of perturbing $\theta$ on the predicted probability of $y$, while the second one captures the impact on all the other classes.

Following this definition, we quantify the importance of $\theta$ with respect to the logic bomb $(\vxs, \vys)$ as $\left|\Delta_{(\vxs, \vys)}^\theta\right|$, which we refer to as the {\em positive impact} of $\theta$.

It seems natural to use a similar definition to measure the importance of $\theta$ with respect to non-target inputs. For example, using the inputs in $\mathcal{R}$ as proxies, we may use $\sum_{(\vx, y) \in \mathcal{R}}\left|\Delta_{(\vx, y)}^\theta\right|$ to measure the overall importance of $\theta$ for non-target inputs. In our empirical study, however, we find that such definitions do not generalize well for inputs not included in $\mathcal{R}$. Instead, inspired by~\cite{Guo:2016:arxiv}, we use the absolute value of the parameter $|\theta|$, a simpler but more effective definition,
 as an indication of its importance for non-target inputs, which we refer to as its {\em negative impact}. This strategy not only improves the crafting efficiency but also reduces the attack's dependency on the accessible training data $\mathcal{R}$.

 %

 \subsubsection*{\bf Attack Model}

 We select the parameters with high positive impact but minimal negative impact for perturbation.
 Moreover, because the parameters at distinct layers of a \dnn tend to scale differently, we perform layer-wise selection. Specifically, we select a candidate parameter if its negative impact value (absolute value) is $\alpha$-standard deviation below the average absolute value of all the parameters at the same layer. Note that by adjusting $\alpha$, we effectively control the number of perturbed parameters (details in \myref{sec:casestudy2}). We collect such candidate parameters across all the layers, and select among them the top $\kappa$ ones with the largest positive impact values to perform the perturbation.

 %


 Putting everything together, Algorithm~\ref{alg:bomb2} sketches our attack model, which iteratively selects a subset of parameters to perturb. At each iteration, for a given layer of the feature extractor, we first compute the threshold of negative impact (line 13); for each parameter $\theta$ at the same layer, we check whether it satisfies the constraint of negative impact and add it to the candidate pool $\bar{\Theta}$ if so (line 14-15); we then pick the top $\kappa$ ones in the pool with the largest positive impact values (line 6-7); for each such parameter $\theta_\sast$, we update $\theta_\sast$ in $\hat{g}$ according to the sign of $\Delta_{(\vxs, \vys)}^{\theta_\sast}$ to increase the likelihood of $\vxs$ being classified as $y_\sast$ (line 8). This process repeats until no more qualified parameters can be found.

\subsection{A Case-Study System}

To empirically validate our attack model, we study a state-of-the-art skin cancer screening system~\cite{Esteva:2017:nature}, which takes as input images of skin lesions and diagnoses potential skin cancers. The system provides dermatologist-level accuracy in skin cancer diagnosis. It was reported that the system achieves $72.1 \pm 0.9\%$ (mean $\pm$ standard deviation) overall accuracy in skin cancer diagnosis; in comparison, two human dermatologists in the study attained $65.56\%$ and $66.0\%$ accuracy.

This case-study system is built upon a DNN-based feature extractor. In particular, it incorporates the feature extractor from the \icep model~\cite{Szegedy:2015:arxiv}, which has been pre-trained on the {\sf ImageNet} dataset~\cite{ILSVRC15}, and performs full-system tuning using the digital skin lesion dataset. The schematic diagram of the case-study system is illustrated in Figure\,\ref{fig:icep}.

\begin{figure}
\centering
\epsfig{file=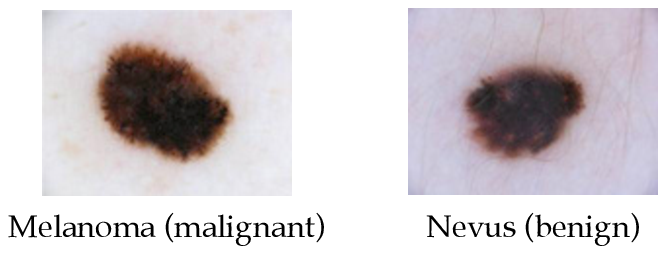, width=65mm}
\caption{Sample skin lesion images of three diseases. \label{fig:image}}
\end{figure}

In their study~\cite{Esteva:2017:nature}, Esteva {\em et al.} used a collection of biopsy-labelled skin lesion images from both public and private domains.  We are able to collect a subset of such images which are publicly available from the International Skin Imaging Collaboration (ISIC) Archive\footnote{ISIC Dermoscopic Archive: https://isic-archive.com}. Similar to~\cite{Esteva:2017:nature}, we categorize these images using a two-disease partition: {\em Melanoma} (malignant) and  {\em Nevus} (benign) lesions, which constitute 708 and 1,633 biospy confirmed images respectively. Figure\,\ref{fig:image} shows one sample image from each category: their similar appearance to human vision demonstrates the difficulty in distinguishing these categories. We split the dataset into 75\% as the training set $\mathcal{T}$ and 25\% as the validation set $\mathcal{V}$. We assume: the system developer uses the entire $\mathcal{T}$ to tune the system; the adversary has access to a reference set $\mathcal{R}$, which a subset of $\mathcal{T}$, and attempts to attack a target patient in  $\mathcal{V}$, but without prior knowledge about other patients in $\mathcal{V}$.
%
%


\subsection{Empirical Evaluation}
\label{sec:casestudy2}

\subsubsection*{\bf Setting} Similar to \myref{sec:attack1}, we first build a baseline system upon genuine feature extractor $g$ and classifier $f$. This baseline system achieves 80.4\% accuracy, which is comparable with~\cite{Esteva:2017:nature}.
 In each trial of the attack, we select an input $\vxs$ sampled from the validation set $\mathcal{V}$ as the target of the adversary. In particular, here we focus on the cases that $\vxs$ truly represents a malignant lesion, while the adversary attempts to force the system to misclassify $\vxs$ as a benign lesion. Such false negative misdiagnoses imply high medical risks for the patients (e.g., preventing them from receiving prompt treatments). We then craft a malicious \mlc $\hat{g}$ to embed the logic bomb $(\vxs, \vys)$, and compare the performance of the infected system $f\circ \hat{g}$ and the baseline system $f\circ g$.

All the models and algorithms are implemented on {\sf TensorFlow}~\cite{tensorflow}, an open source software library for numerical computation using data flow graphs. All the experiments are performed using an array of 4 Nvidia GTX 1080 GPUs. The default setting of the parameters is as follows: the threshold of perturbation amplitude $\epsilon = 2\textrm{e-}3$, the multiple of standard deviation $\alpha = 0.75$, and the fraction of training data accessible by the adversary $|\mathcal{R}|/|\mathcal{T}| = 0.5$. For each set of experiments, we run 156 trials.

\begin{figure}
	\epsfig{file=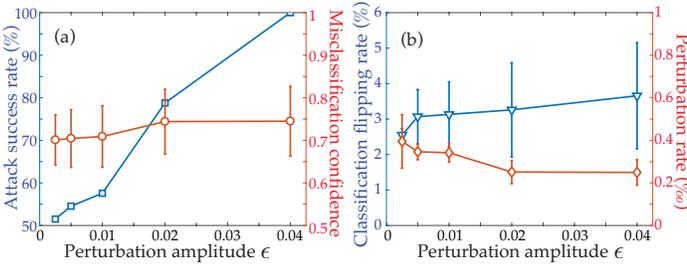, width=92mm}
\caption{Attack effectiveness and evasiveness versus perturbation amplitude $\epsilon$. \label{fig:dnn_succ}}
\end{figure}

\subsubsection*{\bf Summary} Our findings are summarized here.

\begin{myitemize}
\item Effectiveness. We show that logic-bomb attacks are able to trigger the host \ml system to misclassify targeted inputs with 100\%  success rate with average confidence above 0.7, even after the system developer has performed full-system tuning.

\item Evasiveness. We show that the adversary is able to achieve $100\%$ attack success rate via perturbing only 0.03\% of the model parameters, each with distortion less than $10^{-3}$, while the impact on the classification of non-target inputs is below 3.26\%.

\item Easiness. We show that the adversary can readily launch logic-bomb attacks without prior knowledge about the building (e.g., the classifier) of the host \ml system.
\end{myitemize}

Next we detail our studies.

%

%

\subsubsection*{\bf Effectiveness}
\label{sec:effective}

In the first set of experiments, using the two metrics in \myref{sec:attack1exp}, attack success rate and misclassification confidence, we evaluate the effectiveness of logic-bomb attacks against the case-study system.


%
%

%

In Figure\,\ref{fig:dnn_succ} (a), we show the attack's effectiveness as a function of the threshold of perturbation amplitude $\epsilon$ (which controls the perturbation amplitude of individual parameters), where $\epsilon$ varies from $2.5\textrm{e-4}$ to $4\textrm{e-3}$. In Table~\ref{tab:alpha}, we show the attack's effectiveness versus the multiple of standard deviation in parameter selection (Algorithm~\ref{alg:bomb2}), where $\alpha$ varies from 0.5 to 1. Note that in both cases the system is fully tuned.

We have the following observations. First, logic-bomb attack succeeds under fairly low perturbation amplitude. Observe that even with $\epsilon = 2.5\textrm{e-4}$, the adversary is able to force the system to misclassify more than 50\% of the target inputs even under full-system tuning. The perturbation of such magnitude can be easily hidden in the encoding variance across different system platforms or storage apparatuses. Also note that across all the cases, the misclassification confidence remains fairly high (above 70\%). Second, as $\epsilon$ varies from $2.5\textrm{e-4}$ to $4\textrm{e-3}$, the attack success rate grows from 52\% to 100\%, agreeing with our intuition that a larger perturbation amplitude implies more manipulation space for the adversary. Third, the full-system tuning can not fundamentally defend against such attacks. For a reasonably large $\epsilon$ (e.g., $2\textrm{e-3}$), the success rate remains as high as 80\%, which points to the necessity of seeking other more effective defense mechanisms (details in \myref{sec:discussion}). Finally, as $\alpha$ increases from 0.5 to 1.0, both attack success rate and misclassification confidence peak at $\alpha = 0.75$, indicating that a proper setting of $\alpha$ allows to select most effective parameters for perturbation.


%

\begin{table*}{\small

\begin{tabular}{|c|c|c|c|c|}
\hline
\multirow{2}{*}{\bf $\alpha$} &  {\bf Attack} & {\bf Misclassification} & {\bf Classification} &  {\bf Parameter}\\
& {\bf Success Rate (\%)} & {\bf Confidence} & {\bf Flipping Rate (\%)} & {\bf Perturbation Rate (\textperthousand)}\\
\hline
\hline
0.5 & 75.7 & 0.67$\pm$0.07 & 3.34$\pm$1.03 & 0.33$\pm$0.06 \\
0.75 & 99.9 & 0.76$\pm$0.07 & 3.26$\pm$0.91 & 0.34$\pm$0.04 \\
1.0 & 66.7 & 0.70$\pm$0.08 & 3.62$\pm$1.36 & 0.44$\pm$0.05\\
\hline
\end{tabular}}
\caption{Attack effectiveness and evasiveness versus multiple of standard deviation $\alpha$. \label{tab:alpha}}
\end{table*}

\subsubsection*{\bf Evasiveness}
\label{sec:evasive}

In this set of experiments, we evaluate the detection evasiveness of logic-bomb attacks. Since the formulation of Algorithm~\ref{alg:bomb2} directly bounds the perturbation amplitude on individual parameters, we use the metric of {\em parameter perturbation rate},  which is the fraction of perturbed parameters, to quantify the syntactic indiscernibility, and use the metric of classification flipping rate (\myref{sec:attack1exp}) to quantify the semantic indiscernibility.

Figure\,\ref{fig:dnn_succ} (b) shows the attack's evasiness as a function of the threshold of perturbation amplitude $\epsilon$, and Table~\ref{tab:alpha} shows the attack's evasiveness versus the multiple of standard deviation in parameter selection (Algorithm~\ref{alg:bomb2}).

We have the following observations. First, the classification flipping rate is positively correlated with the allowable perturbation amplitude $\epsilon$, which however grow fairly slowly. For example, as $\epsilon$ increases from $2.5\textrm{e-4}$ to $4\textrm{e-3}$, the average flipping rate grows less than 1\%. Second, the number of perturbed parameters is extremely low (e.g., less than 0.33\%), compared with the total number of parameters in the DNN. Third, a larger perturbation amplitude entails a lower parameter perturbation rate (i.e., less number of parameters to be perturbed). Finally, the parameter perturbation rate also decreases as the multiple of standard deviation $\alpha$ grows. This is explained by that a larger $\alpha$ means a stricter parameter selection criterion, leading to lower number of candidate parameters. Overall, we can conclude that under proper parameter settings, logic-bomb attack incur fairly indiscernible impact on non-target inputs.


Astute readers may notice that the classification flipping rate here is relatively higher than the cases in \myref{sec:attack1exp}. This is partially attributed to the inherent randomness in DNN training (e.g., random initialization, dropout layers, stochastic gradient descent), each time training the same DNN model on the same training set may result in a slightly different model.

\subsection{Easiness}
\label{sec:resource}
%
%
%

\begin{figure}
\centering
\epsfig{file=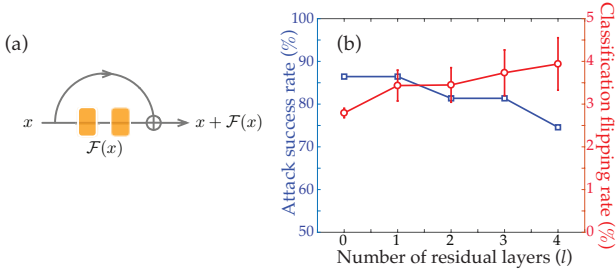, width=84mm}
\caption{Impact of the instantiation of classifier $f$ on the attack's effectiveness and evasiveness. \label{fig:classifier2}}
\end{figure}

Finally, we assess the impact of the adversary's knowledge about the classifier in the host \ml system. Specifically,
we consider variants of the basic classifier $f$ shown in Figure\,\ref{fig:icep} by appending a set of residual layers at its input end. As shown in Figure\,\ref{fig:classifier2}(a), with reference to the layer input $x$, a residual layer is designed to learn a residual function $\mathcal{F}(\cdot)$, which often helps adapt the model learned in another domain to the current domain~\cite{He:2015:arxiv}. By varying the number of residual layers ($l$) appended to $f$, we create a set of variants. It is clear that the classifier becomes increasingly different from the adversary's surrogate classifier as $l$ grows.

Observe in Figure\,\ref{fig:classifier2} that the attack's effectiveness (attack success rate) is fairly insensitive to the parameter $l$. For example, as $l$ increases from 0 to 4 (0 corresponds to the original classifier), the attack success rate varies about 10\%. This validates our analysis in \myref{sec:attack1} that the concrete form of the surrogate classifier is often immaterial. Meanwhile, the attack's evasiveness is insensitive to $l$ as well. The average classification flipping rate gradually grows from about 2.7\% to less than 4\%, as $l$ increases from 0 to 4.



\section{Discussion}
\label{sec:discussion}


In \myref{sec:attack1} and \myref{sec:attack2}, we have empirically demonstrated that, with fairly indiscernible distortion, it is feasible to craft malicious \mlcs that trigger host \ml systems to malfunction in a predictable manner. Here we provide analytical justification for the success of such attacks by relating to some recent theoretical advances. Based on this analysis, we further discuss potential countermeasures to mitigate \mlc-based attacks.

\begin{table*}{\small

\begin{tabular}{|c|c|c|c|c|c|c|c|c|c|}

\hline
\multirow{2}{*}{\bf Classifier} & \multirow{2}{*}{\bf Disease} &  \multicolumn{4}{c}{\bf Attack Success Rate (\%)} & \multicolumn{4}{|c|}{\bf Classification Flipping Rate (\%)}\\
\cline{3-10}
& & $D = 50$ & $D = 100$ & $D = 150$ & $D = 200$  & $D = 50$ & $D = 100$ & $D = 150$ & $D = 200$ \\
\hline
\hline

\multirow{4}{*}{\bf LR} & {\em d\_427} & 75.8 & 78.8 & 78.8 & 81.8 & 0.79$\pm$0.17 & 0.69$\pm$0.10 & 0.57$\pm$0.15 & 0.50$\pm$0.17 \\
 & {\em d\_428} & 72.7 & 75.8 & 78.8 & 78.8 & 1.08$\pm$0.25 & 0.67$\pm$0.14 & 0.57$\pm$0.13 & 0.53$\pm$0.16 \\
 & {\em d\_250} & 69.7 & 78.8 & 81.8 & 81.8 & 1.18$\pm$0.21 & 1.09$\pm$0.14 & 0.84$\pm$0.17 & 0.52$\pm$0.14 \\
 & {\em d\_272} & 87.9 & 93.9 & 93.9 & 97.0 & 0.89$\pm$0.20 & 0.65$\pm$0.20 & 0.56$\pm$0.19 & 0.39$\pm$0.14 \\
\hline
\hline
\multirow{4}{*}{\bf SVM} & {\em d\_427} & 78.8 & 78.8 & 78.8 & 81.8 & 0.76$\pm$0.17 & 0.72$\pm$0.09 & 0.66$\pm$0.15 & 0.39$\pm$0.14 \\
 & {\em d\_428} & 72.7 & 78.8 & 78.8 & 81.8 & 1.12$\pm$0.28 & 0.70$\pm$0.11 & 0.64$\pm$0.13 & 0.32$\pm$0.12 \\
 & {\em d\_250} & 60.6 & 72.7 & 75.8 & 78.8 & 1.16$\pm$0.20 & 0.91$\pm$0.14 & 0.79$\pm$0.17 & 0.41$\pm$0.12 \\
 & {\em d\_272} & 90.9 & 90.9 & 93.9 & 93.9 & 0.59$\pm$0.21 & 0.56$\pm$0.16 & 0.46$\pm$0.17 & 0.44$\pm$0.16 \\
\hline
\hline
\multirow{4}{*}{\bf MLP} & {\em d\_427} & 60.6 & 63.6 & 66.7 & 72.7 & 1.03$\pm$0.08 & 0.95$\pm$0.14 & 0.77$\pm$0.29 & 0.61$\pm$0.29 \\
 & {\em d\_428} & 63.6 & 63.6 & 63.6 & 69.7 & 1.36$\pm$0.24 & 1.33$\pm$0.13 & 1.13$\pm$0.16 & 0.77$\pm$0.21 \\
 & {\em d\_250} & 48.5 & 57.6 & 72.7 & 75.8 & 1.26$\pm$0.06 & 1.22$\pm$0.16 & 0.45$\pm$0.08 & 0.38$\pm$0.20 \\
 & {\em d\_272} & 63.6 & 72.7 & 84.8 & 87.9 & 1.28$\pm$0.18 & 1.21$\pm$0.09 & 0.53$\pm$0.28 & 0.43$\pm$0.26 \\
\hline

\end{tabular}}

\caption{Impact of model complexity on attack's effectiveness and evasiveness. \label{tab:complexity}}

\end{table*}

\subsection{Why do logic-bomb attacks work?}
\label{sec:success}

Many of today's \mlcs are complex \ml models designed to model highly non-linear, non-convex functions. For instance, in the case of DNN-based \mlc, according to the universal approximation theorem~\cite{Hornik:1991:nn}, a feed-forward neural network with only a single hidden layer is capable of describing any continuous functions. Recent studies~\cite{Zhang:2016:arxiv} have further provided both empirical and theoretical evidence that the effective capacity of many \dnns is sufficient for ``memorizing'' the entire training set.

\begin{figure}
\centering
\epsfig{file=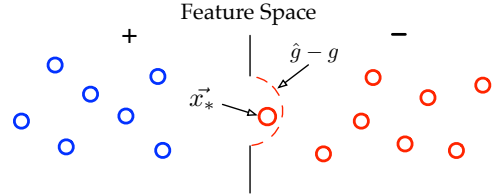, width=65mm}
\caption{Alteration of the underlying distribution of feature vectors by the logic-bomb attack. \label{fig:change}}
\end{figure}

The observations above may partially explain the phenomenon that under carefully perturbation, an \mlc is able to memorize a singular input (i.e., the target victim) yet without comprising its generalization to other non-target inputs. This phenomenon is illustrated in Figure\,\ref{fig:change}. Intuitively, in the manifold space spanned by all the feature vectors, the perturbation  $(\hat{g} - g)$ alters the boundaries between different classes in order to change the classification of $\vxs$; yet, thanks to the complexity of the \ml model, this alteration is performed in such a precise manner that only the proximate space of $\vxs$ is affected, without noticeable influence to other inputs.

To verify the analysis, we empirically assess the impact of the \mlc model complexity on the attack's effectiveness and evasiveness. Apparently, under the  setting of \myref{sec:attack1exp}, the feature space dimensionality (i.e., the embedding vector dimensionality) directly determines the \mlc model complexity. We thus measure the attack's success rate and classification flipping rate under varying feature space dimensionalities.

The results are shown in Table\,\ref{tab:complexity}. Observe that the increasing model complexity benefits both effectiveness and evasiveness: as the feature space dimensionality varies from 50 to 200, across all the cases (classifiers + disease), the attack success rate increases, while the classification flipping rate decreases. For example, in the case of MLP and d\_427, the success rate grows by 12.1\% while the flipping rate drops by 0.42\%. It is thus reasonable to postulate the existence of strong connections between the complexity of \mlc models and the success of logic-bomb attacks.

\subsection{How can logic-bomb attacks be defended?}

The \ml system developers now face a dilemma. On the one hand, the ever-increasing system complexity makes \mlc-based development not only tempting but also necessary; on the other hand, the potential risks of third-party \mlcs may significantly undermine the safety of \ml systems in security-critical domains. Below we discuss a few possible defense strategies based on the sources of \mlcs and discuss their associated technical challenges.

For \mlcs contributed by reputable sources (e.g., Google Brain), the primary task is to verify the authenticity of \mlcs. The digital signature machinery may seem one straightforward solution, which however entails non-trivial challenges. The first one is its efficiency. Many \mlcs (e.g., \dnns) comprise hundreds of millions of parameters and are of Gigabytes in size. The second one is the encoding variance. Storing and transferring \mlcs across different platforms (e.g., 16-bit versus 32-bit floating numbers) results in fairly different models, while, as shown in \myref{sec:attack1} and \myref{sec:attack2}, even a slight difference of $10^{-4}$ allows the adversary to successfully launch the attacks. To address this issue, it may be necessary for the contributors to publish platform-specific \mlcs.

For \mlcs contributed by untrusted sources, the primary task is to vet the integrity of \mlcs for potential logic bombs. As shown in Figure\,\ref{fig:change}, this amounts to searching for irregular boundaries induced by the \mlc in the feature space. However, it is infeasible to run exhaustive search due to its high dimensionality. A more feasible strategy is to perform anomaly detection based on the training set. Specifically, if a feature extractor \mlc generates a vastly different feature vector for a particular input among a group of similar inputs, this specific input may be proximate to a potential logic bomb, which warrants for further investigation. Clearly, this solution requires that the training set is sufficiently representative for possible inputs encountered during the inference time, which nevertheless may not hold in real settings.

\begin{figure}
	\centering
	\epsfig{file=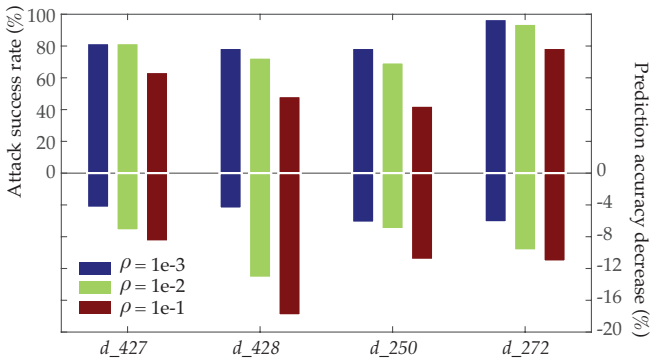, width=86mm}
\caption{Impact of noise addition on attack success rate and classification accuracy. \label{fig:defense}}
\end{figure}

One may also suggest to inject noise to a suspicious \mlc to counter the potential manipulation. We conduct an experiment to show the  challenge of this method. Under the default setting of \myref{sec:attack1exp}, we add random noise sampled from a uniform distribution $[-\rho, \rho]$ to each element of the word embedding. We measure the attack success rate and the classification accuracy decrease versus the setting of $\rho$. As shown in Figure\,\ref{fig:defense}, indeed as $\rho$ increases, the attack is mitigated to a certain extent, which however is achieved at the expense of classification accuracy. In the case of d\_428, the noise of $\rho = 1\textrm{e-}1$ incurs as much as 17\% accuracy drop. Apparently, a delicate balance needs to be struck.

Besides the logic-bomb attacks, we envision that \mlcs may also function as vehicles for other attacks against \ml systems (e.g., model inversion attacks~\cite{Fredrikson:2015:ccs} and model extraction attacks~\cite{Tramer:2016:sec}), which apparently require different countermeasures.

\section{Related Work}
\label{sec:literature}

Next we review three categories of related work: adversarial machine learning, deep learning-specific attacks, and external software library-based attacks.

Lying at the core of many security-critical domains, \ml systems are increasingly becoming the targets of malicious attacks~\cite{Barreno:2006:asiaccs,Huang:2011:aisec,Barreno:2010:SML}. Two primary threat models are considered in literature. (i) Poisoning attacks, in which the adversary pollutes the training data to eventually compromise the  \ml systems~\cite{Biggio:2012:icml,Xiao:2015:SVM,Rubinstein:2009:imc}. (ii) Evasion attacks, in which the adversary modifies the input data during the inference time to trigger the systems to misbehave~\cite{Dalvi:2004:kdd,Lowd:2005:kdd,Nelson:2012:QSE}. To our best knowledge, this work is among the first few to investigate \mlc-based attacks, in which the adversary leverages compromised \mlcs to influence the system behaviors.

Compared with simple \ml models (e.g., decision tree, support vector machine, logistic regression), securing deep learning systems deployed in adversarial settings poses even more challenges for they are designed to model highly nonlinear, nonconvex functions~\cite{LeCun:2015:nature}. One line of work focuses on developing new attack strategies against \dnns~\cite{Goodfellow:2014:arxiv,Huang:2015:arxiv,Tabacof:2015:arXiv,Papernot:2016:eurosp,Carlini:2017:sp}, attempting to find the minimum possible distortion to the input data to trigger the systems to misclassify. Another line of work strives to improve \dnn resilience against such adversarial input attacks by developing new training and inference regimes~\cite{Goodfellow:2014:arxiv,Gu:2014:arxiv,Huang:2015:arxiv,Papernot:2016:sp}. However, none of the work has considered exploiting DNN-based \mlcs to compromise \ml systems, not to mention mitigating such threats.


Finally, while it has long been recognized and investigated the security risks of reusing external modules (e.g., libraries) in software development (e.g.,~\cite{libpng,heartbleed}), especially in web and mobile applications~\cite{Roesner:2013:sec,Bhoraskar:2014:sec,Chen:2016:sp}, it is still challenging today even to reliably detect external modules~\cite{Backes:2016:ccs}, because of the ever-increasing system complexity. Addressing the security risks of external modules in building and operating \ml systems presents even more challenges, due to their ``stateful'' nature (i.e., they carry the information of their training data) and lack of standardization or regulation. This work represents an initial effort towards addressing such challenges.


%
%
%
%

\section{Conclusion}
\label{sec:conclusion}


This work represents an in-depth study on the security implications of using third-party \mlcs in building and operating \ml systems. Exemplifying with two real \ml systems in the healthcare domain, we demonstrated a broad class of logic-bomb attacks that trigger host \ml systems to malfunction in a predictable manner. We provided analytical justification for the success of such attacks, which points to the fundamental characteristics of today's \ml models: high dimensionality, non-linearity, and non-convexity. Thus, this issue seems fundamental to many \ml systems.

It is our hope that this work can raise the awareness of the security and \ml research communities about this important issue. A few possible avenues for further investigation include: First, besides logic-bomb attacks presented in this paper, it is interesting to explore \mlcs as vehicles to launch other types of attacks (e.g., facilitating to extract sensitive information about input data). Second, this paper only considered the attacks based on a single \mlc. We speculate that the attacks leveraging multiple colluding \mlcs would be even more damaging and detection-evasive. Finally, implementing and evaluating the countermeasures proposed in \myref{sec:discussion} in real \ml systems may serve as a promising starting point for developing effective defense mechanisms.

\newpage

\bibliographystyle{ACM-Reference-Format}
\bibliography{modular}


\begin{thebibliography}{00}


\ifx \showCODEN    \undefined \def \showCODEN     #1{\unskip}     \fi
\ifx \showDOI      \undefined \def \showDOI       #1{#1}\fi
\ifx \showISBNx    \undefined \def \showISBNx     #1{\unskip}     \fi
\ifx \showISBNxiii \undefined \def \showISBNxiii  #1{\unskip}     \fi
\ifx \showISSN     \undefined \def \showISSN      #1{\unskip}     \fi
\ifx \showLCCN     \undefined \def \showLCCN      #1{\unskip}     \fi
\ifx \shownote     \undefined \def \shownote      #1{#1}          \fi
\ifx \showarticletitle \undefined \def \showarticletitle #1{#1}   \fi
\ifx \showURL      \undefined \def \showURL       {\relax}        \fi
\providecommand\bibfield[2]{#2}
\providecommand\bibinfo[2]{#2}
\providecommand\natexlab[1]{#1}
\providecommand\showeprint[2][]{arXiv:#2}

\bibitem[\protect\citeauthoryear{Ahn}{Ahn}{2016}]%
        {where2go}
\bibfield{author}{\bibinfo{person}{Dong~Won Ahn}.}
  \bibinfo{year}{2016}\natexlab{}.
\newblock \bibinfo{title}{Where2go: travel destination recommendation}.
\newblock \bibinfo{howpublished}{\url{http://where2go.help}}.
  (\bibinfo{year}{2016}).
\newblock


\bibitem[\protect\citeauthoryear{An open-source software library for Machine
  Intelligence}{An open-source software library for Machine
  Intelligence}{2015}]%
        {tensorflow}
An open-source software library for Machine Intelligence
  \bibinfo{year}{2015}\natexlab{}.
\newblock \bibinfo{howpublished}{\url{https://www.tensorflow.org}}.
  (\bibinfo{year}{2015}).
\newblock


\bibitem[\protect\citeauthoryear{Asgari and Mofrad}{Asgari and Mofrad}{2015}]%
        {Asgari:2015:plos}
\bibfield{author}{\bibinfo{person}{Ehsaneddin Asgari} {and}
  \bibinfo{person}{Mohammad R.~K. Mofrad}.} \bibinfo{year}{2015}\natexlab{}.
\newblock \showarticletitle{Continuous Distributed Representation of Biological
  Sequences for Deep Proteomics and Genomics}.
\newblock \bibinfo{journal}{{\em PLOS ONE\/}} \bibinfo{volume}{10},
  \bibinfo{number}{11} (\bibinfo{date}{11} \bibinfo{year}{2015}),
  \bibinfo{pages}{1--15}.
\newblock


\bibitem[\protect\citeauthoryear{Backes, Bugiel, and Derr}{Backes
  et~al\mbox{.}}{2016}]%
        {Backes:2016:ccs}
\bibfield{author}{\bibinfo{person}{Michael Backes}, \bibinfo{person}{Sven
  Bugiel}, {and} \bibinfo{person}{Erik Derr}.} \bibinfo{year}{2016}\natexlab{}.
\newblock \showarticletitle{Reliable Third-Party Library Detection in Android
  and Its Security Applications}. In \bibinfo{booktitle}{{\em Proceedings of
  the 2016 ACM SIGSAC Conference on Computer and Communications Security}} {\em
  (\bibinfo{series}{CCS '16})}.
\newblock


\bibitem[\protect\citeauthoryear{Barreno, Nelson, Joseph, and Tygar}{Barreno
  et~al\mbox{.}}{2010}]%
        {Barreno:2010:SML}
\bibfield{author}{\bibinfo{person}{Marco Barreno}, \bibinfo{person}{Blaine
  Nelson}, \bibinfo{person}{Anthony~D. Joseph}, {and} \bibinfo{person}{J.~D.
  Tygar}.} \bibinfo{year}{2010}\natexlab{}.
\newblock \showarticletitle{The Security of Machine Learning}.
\newblock \bibinfo{journal}{{\em Mach. Learn.\/}} \bibinfo{volume}{81},
  \bibinfo{number}{2} (\bibinfo{year}{2010}), \bibinfo{pages}{121--148}.
\newblock


\bibitem[\protect\citeauthoryear{Barreno, Nelson, Sears, Joseph, and
  Tygar}{Barreno et~al\mbox{.}}{2006}]%
        {Barreno:2006:asiaccs}
\bibfield{author}{\bibinfo{person}{Marco Barreno}, \bibinfo{person}{Blaine
  Nelson}, \bibinfo{person}{Russell Sears}, \bibinfo{person}{Anthony~D.
  Joseph}, {and} \bibinfo{person}{J.~D. Tygar}.}
  \bibinfo{year}{2006}\natexlab{}.
\newblock \showarticletitle{Can Machine Learning Be Secure?}. In
  \bibinfo{booktitle}{{\em ASIACCS}}.
\newblock


\bibitem[\protect\citeauthoryear{Bhoraskar, Han, Jeon, Azim, Chen, Jung, Nath,
  Wang, and Wetherall}{Bhoraskar et~al\mbox{.}}{2014}]%
        {Bhoraskar:2014:sec}
\bibfield{author}{\bibinfo{person}{Ravi Bhoraskar}, \bibinfo{person}{Seungyeop
  Han}, \bibinfo{person}{Jinseong Jeon}, \bibinfo{person}{Tanzirul Azim},
  \bibinfo{person}{Shuo Chen}, \bibinfo{person}{Jaeyeon Jung},
  \bibinfo{person}{Suman Nath}, \bibinfo{person}{Rui Wang}, {and}
  \bibinfo{person}{David Wetherall}.} \bibinfo{year}{2014}\natexlab{}.
\newblock \showarticletitle{Brahmastra: Driving Apps to Test the Security of
  Third-party Components}. In \bibinfo{booktitle}{{\em Proceedings of the 23rd
  USENIX Conference on Security Symposium}} {\em (\bibinfo{series}{SEC'14})}.
\newblock


\bibitem[\protect\citeauthoryear{Biggio, Nelson, and Laskov}{Biggio
  et~al\mbox{.}}{2012}]%
        {Biggio:2012:icml}
\bibfield{author}{\bibinfo{person}{Battista Biggio}, \bibinfo{person}{Blaine
  Nelson}, {and} \bibinfo{person}{Pavel Laskov}.}
  \bibinfo{year}{2012}\natexlab{}.
\newblock \showarticletitle{Poisoning Attacks against Support Vector Machines}.
  In \bibinfo{booktitle}{{\em ICML}}.
\newblock


\bibitem[\protect\citeauthoryear{{BVLC}}{{BVLC}}{2017}]%
        {modelzoo}
\bibfield{author}{\bibinfo{person}{{BVLC}}.} \bibinfo{year}{2017}\natexlab{}.
\newblock \bibinfo{title}{Model Zoo}.
\newblock
  \bibinfo{howpublished}{\url{https://github.com/BVLC/caffe/wiki/Model-Zoo}}.
  (\bibinfo{year}{2017}).
\newblock


\bibitem[\protect\citeauthoryear{{Carlini} and {Wagner}}{{Carlini} and
  {Wagner}}{2017}]%
        {Carlini:2017:sp}
\bibfield{author}{\bibinfo{person}{N. {Carlini}} {and} \bibinfo{person}{D.
  {Wagner}}.} \bibinfo{year}{2017}\natexlab{}.
\newblock \showarticletitle{{Towards Evaluating the Robustness of Neural
  Networks}}. In \bibinfo{booktitle}{{\em Proceedings of the 38th IEEE
  Symposium on Security and Privacy}} {\em (\bibinfo{series}{S\&P '17})}.
\newblock


\bibitem[\protect\citeauthoryear{Chen, Wang, Chen, Wang, Lee, Wang, Ma, Wang,
  Zhang, and Zou}{Chen et~al\mbox{.}}{2016}]%
        {Chen:2016:sp}
\bibfield{author}{\bibinfo{person}{K. Chen}, \bibinfo{person}{X. Wang},
  \bibinfo{person}{Y. Chen}, \bibinfo{person}{P. Wang}, \bibinfo{person}{Y.
  Lee}, \bibinfo{person}{X. Wang}, \bibinfo{person}{B. Ma}, \bibinfo{person}{A.
  Wang}, \bibinfo{person}{Y. Zhang}, {and} \bibinfo{person}{W. Zou}.}
  \bibinfo{year}{2016}\natexlab{}.
\newblock \showarticletitle{Following Devil's Footprints: Cross-Platform
  Analysis of Potentially Harmful Libraries on Android and iOS}. In
  \bibinfo{booktitle}{{\em 2016 IEEE Symposium on Security and Privacy (SP)}}.
\newblock


\bibitem[\protect\citeauthoryear{Dalvi, Domingos, Mausam, Sanghai, and
  Verma}{Dalvi et~al\mbox{.}}{2004}]%
        {Dalvi:2004:kdd}
\bibfield{author}{\bibinfo{person}{Nilesh Dalvi}, \bibinfo{person}{Pedro
  Domingos}, \bibinfo{person}{Mausam}, \bibinfo{person}{Sumit Sanghai}, {and}
  \bibinfo{person}{Deepak Verma}.} \bibinfo{year}{2004}\natexlab{}.
\newblock \showarticletitle{Adversarial Classification}. In
  \bibinfo{booktitle}{{\em KDD}}.
\newblock


\bibitem[\protect\citeauthoryear{Esteva, Kuprel, Novoa, Ko, Swetter, Blau, and
  Thrun}{Esteva et~al\mbox{.}}{2017}]%
        {Esteva:2017:nature}
\bibfield{author}{\bibinfo{person}{Andre Esteva}, \bibinfo{person}{Brett
  Kuprel}, \bibinfo{person}{Roberto~A. Novoa}, \bibinfo{person}{Justin Ko},
  \bibinfo{person}{Susan~M. Swetter}, \bibinfo{person}{Helen~M. Blau}, {and}
  \bibinfo{person}{Sebastian Thrun}.} \bibinfo{year}{2017}\natexlab{}.
\newblock \showarticletitle{Dermatologist-level classification of skin cancer
  with deep neural networks}.
\newblock \bibinfo{journal}{{\em Nature\/}} \bibinfo{volume}{542},
  \bibinfo{number}{7639} (\bibinfo{year}{2017}), \bibinfo{pages}{115--118}.
\newblock


\bibitem[\protect\citeauthoryear{Farhan, Wang, Huang, Wang, Wang, and
  Jiang}{Farhan et~al\mbox{.}}{2016}]%
        {farhan:2016:jmir}
\bibfield{author}{\bibinfo{person}{Wael Farhan}, \bibinfo{person}{Zhimu Wang},
  \bibinfo{person}{Yingxiang Huang}, \bibinfo{person}{Shuang Wang},
  \bibinfo{person}{Fei Wang}, {and} \bibinfo{person}{Xiaoqian Jiang}.}
  \bibinfo{year}{2016}\natexlab{}.
\newblock \showarticletitle{A Predictive Model for Medical Events Based on
  Contextual Embedding of Temporal Sequences}.
\newblock \bibinfo{journal}{{\em JMIR Med Inform\/}} \bibinfo{volume}{4},
  \bibinfo{number}{4} (\bibinfo{year}{2016}), \bibinfo{pages}{e39}.
\newblock


\bibitem[\protect\citeauthoryear{Fratantonio, Bianchi, Robertson, Kirda,
  Kruegel, and Vigna}{Fratantonio et~al\mbox{.}}{2016}]%
        {Fratantonio:2016:sp}
\bibfield{author}{\bibinfo{person}{Yanick Fratantonio},
  \bibinfo{person}{Antonio Bianchi}, \bibinfo{person}{William Robertson},
  \bibinfo{person}{Engin Kirda}, \bibinfo{person}{Christopher Kruegel}, {and}
  \bibinfo{person}{Giovanni Vigna}.} \bibinfo{year}{2016}\natexlab{}.
\newblock \showarticletitle{TriggerScope: Towards Detecting Logic Bombs in
  Android Applications}. In \bibinfo{booktitle}{{\em Proceedings of the 2016
  IEEE Symposium on Security and Privacy (SP)}} {\em (\bibinfo{series}{S\&P
  '16})}.
\newblock


\bibitem[\protect\citeauthoryear{Fredrikson, Jha, and Ristenpart}{Fredrikson
  et~al\mbox{.}}{2015}]%
        {Fredrikson:2015:ccs}
\bibfield{author}{\bibinfo{person}{Matt Fredrikson}, \bibinfo{person}{Somesh
  Jha}, {and} \bibinfo{person}{Thomas Ristenpart}.}
  \bibinfo{year}{2015}\natexlab{}.
\newblock \showarticletitle{Model Inversion Attacks That Exploit Confidence
  Information and Basic Countermeasures}. In \bibinfo{booktitle}{{\em
  Proceedings of the 22Nd ACM SIGSAC Conference on Computer and Communications
  Security}} {\em (\bibinfo{series}{CCS '15})}.
\newblock


\bibitem[\protect\citeauthoryear{Garling}{Garling}{2014}]%
        {sfgate}
\bibfield{author}{\bibinfo{person}{Caleb Garling}.}
  \bibinfo{year}{2014}\natexlab{}.
\newblock \bibinfo{title}{ThisPlusThat.me tries to decode human language}.
\newblock
  \bibinfo{howpublished}{\url{http://www.sfgate.com/technology/article/}}.
  (\bibinfo{year}{2014}).
\newblock


\bibitem[\protect\citeauthoryear{GitHub: The world's leading software
  development platform}{GitHub: The world's leading software development
  platform}{2008}]%
        {github}
GitHub: The world's leading software development platform
  \bibinfo{year}{2008}\natexlab{}.
\newblock \bibinfo{howpublished}{\url{https://github.com}}.
  (\bibinfo{year}{2008}).
\newblock


\bibitem[\protect\citeauthoryear{Globerson, Chechik, Pereira, and
  Tishby}{Globerson et~al\mbox{.}}{2007}]%
        {Globerson:2007:jmlr}
\bibfield{author}{\bibinfo{person}{Amir Globerson}, \bibinfo{person}{Gal
  Chechik}, \bibinfo{person}{Fernando Pereira}, {and} \bibinfo{person}{Naftali
  Tishby}.} \bibinfo{year}{2007}\natexlab{}.
\newblock \showarticletitle{Euclidean Embedding of Co-occurrence Data}.
\newblock \bibinfo{journal}{{\em J. Mach. Learn. Res.\/}}  \bibinfo{volume}{8}
  (\bibinfo{year}{2007}), \bibinfo{pages}{2265--2295}.
\newblock
\showISSN{1532-4435}


\bibitem[\protect\citeauthoryear{{Goodfellow}, {Shlens}, and
  {Szegedy}}{{Goodfellow} et~al\mbox{.}}{2014}]%
        {Goodfellow:2014:arxiv}
\bibfield{author}{\bibinfo{person}{I.~J. {Goodfellow}}, \bibinfo{person}{J.
  {Shlens}}, {and} \bibinfo{person}{C. {Szegedy}}.}
  \bibinfo{year}{2014}\natexlab{}.
\newblock \showarticletitle{{Explaining and Harnessing Adversarial Examples}}.
\newblock \bibinfo{journal}{{\em ArXiv e-prints\/}} (\bibinfo{year}{2014}).
\newblock


\bibitem[\protect\citeauthoryear{{Grosse}, {Papernot}, {Manoharan}, {Backes},
  and {McDaniel}}{{Grosse} et~al\mbox{.}}{2016}]%
        {Grosse:arxiv:2016}
\bibfield{author}{\bibinfo{person}{K. {Grosse}}, \bibinfo{person}{N.
  {Papernot}}, \bibinfo{person}{P. {Manoharan}}, \bibinfo{person}{M. {Backes}},
  {and} \bibinfo{person}{P. {McDaniel}}.} \bibinfo{year}{2016}\natexlab{}.
\newblock \showarticletitle{{Adversarial Perturbations Against Deep Neural
  Networks for Malware Classification}}.
\newblock \bibinfo{journal}{{\em ArXiv e-prints\/}} (\bibinfo{year}{2016}).
\newblock


\bibitem[\protect\citeauthoryear{{Gu} and {Rigazio}}{{Gu} and
  {Rigazio}}{2014}]%
        {Gu:2014:arxiv}
\bibfield{author}{\bibinfo{person}{S. {Gu}} {and} \bibinfo{person}{L.
  {Rigazio}}.} \bibinfo{year}{2014}\natexlab{}.
\newblock \showarticletitle{{Towards Deep Neural Network Architectures Robust
  to Adversarial Examples}}.
\newblock \bibinfo{journal}{{\em ArXiv e-prints\/}} (\bibinfo{year}{2014}).
\newblock


\bibitem[\protect\citeauthoryear{{Guo}, {Yao}, and {Chen}}{{Guo}
  et~al\mbox{.}}{2016}]%
        {Guo:2016:arxiv}
\bibfield{author}{\bibinfo{person}{Y. {Guo}}, \bibinfo{person}{A. {Yao}}, {and}
  \bibinfo{person}{Y. {Chen}}.} \bibinfo{year}{2016}\natexlab{}.
\newblock \showarticletitle{{Dynamic Network Surgery for Efficient DNNs}}.
\newblock \bibinfo{journal}{{\em ArXiv e-prints\/}} (\bibinfo{year}{2016}).
\newblock


\bibitem[\protect\citeauthoryear{He, Zhang, Ren, and Sun}{He
  et~al\mbox{.}}{2015}]%
        {He:2015:arxiv}
\bibfield{author}{\bibinfo{person}{Kaiming He}, \bibinfo{person}{Xiangyu
  Zhang}, \bibinfo{person}{Shaoqing Ren}, {and} \bibinfo{person}{Jian Sun}.}
  \bibinfo{year}{2015}\natexlab{}.
\newblock \showarticletitle{Deep Residual Learning for Image Recognition}.
\newblock \bibinfo{journal}{{\em ArXiv e-prints\/}} (\bibinfo{year}{2015}).
\newblock


\bibitem[\protect\citeauthoryear{Hornik}{Hornik}{1991}]%
        {Hornik:1991:nn}
\bibfield{author}{\bibinfo{person}{Kurt Hornik}.}
  \bibinfo{year}{1991}\natexlab{}.
\newblock \showarticletitle{Approximation Capabilities of Multilayer
  Feedforward Networks}.
\newblock \bibinfo{journal}{{\em Neural Netw.\/}} \bibinfo{volume}{4},
  \bibinfo{number}{2} (\bibinfo{year}{1991}), \bibinfo{pages}{251--257}.
\newblock


\bibitem[\protect\citeauthoryear{Huang, Joseph, Nelson, Rubinstein, and
  Tygar}{Huang et~al\mbox{.}}{2011}]%
        {Huang:2011:aisec}
\bibfield{author}{\bibinfo{person}{Ling Huang}, \bibinfo{person}{Anthony~D.
  Joseph}, \bibinfo{person}{Blaine Nelson}, \bibinfo{person}{Benjamin~I.P.
  Rubinstein}, {and} \bibinfo{person}{J.~D. Tygar}.}
  \bibinfo{year}{2011}\natexlab{}.
\newblock \showarticletitle{Adversarial Machine Learning}. In
  \bibinfo{booktitle}{{\em AISec}}.
\newblock


\bibitem[\protect\citeauthoryear{{Huang}, {Xu}, {Schuurmans}, and
  {Szepesvari}}{{Huang} et~al\mbox{.}}{2015}]%
        {Huang:2015:arxiv}
\bibfield{author}{\bibinfo{person}{R. {Huang}}, \bibinfo{person}{B. {Xu}},
  \bibinfo{person}{D. {Schuurmans}}, {and} \bibinfo{person}{C. {Szepesvari}}.}
  \bibinfo{year}{2015}\natexlab{}.
\newblock \showarticletitle{{Learning with a Strong Adversary}}.
\newblock \bibinfo{journal}{{\em ArXiv e-prints\/}} (\bibinfo{year}{2015}).
\newblock


\bibitem[\protect\citeauthoryear{Johnson, Pollard, Shen, Lehman, Feng,
  Ghassemi, Moody, Szolovits, Anthony~Celi, and Mark}{Johnson
  et~al\mbox{.}}{2016}]%
        {mimic}
\bibfield{author}{\bibinfo{person}{Alistair E.~W. Johnson},
  \bibinfo{person}{Tom~J. Pollard}, \bibinfo{person}{Lu Shen},
  \bibinfo{person}{Li-wei~H. Lehman}, \bibinfo{person}{Mengling Feng},
  \bibinfo{person}{Mohammad Ghassemi}, \bibinfo{person}{Benjamin Moody},
  \bibinfo{person}{Peter Szolovits}, \bibinfo{person}{Leo Anthony~Celi}, {and}
  \bibinfo{person}{Roger~G. Mark}.} \bibinfo{year}{2016}\natexlab{}.
\newblock \showarticletitle{MIMIC-III, a freely accessible critical care
  database}.
\newblock \bibinfo{journal}{{\em Scientific Data\/}}  \bibinfo{volume}{3}
  (\bibinfo{year}{2016}).
\newblock


\bibitem[\protect\citeauthoryear{Kepes}{Kepes}{2015}]%
        {ml-legal}
\bibfield{author}{\bibinfo{person}{Ben Kepes}.}
  \bibinfo{year}{2015}\natexlab{}.
\newblock \bibinfo{title}{eBrevia Applies Machine Learning To Contract Review}.
\newblock \bibinfo{howpublished}{\url{https://www.forbes.com/}}.
  (\bibinfo{year}{2015}).
\newblock


\bibitem[\protect\citeauthoryear{Krizhevsky, Sutskever, and Hinton}{Krizhevsky
  et~al\mbox{.}}{2012}]%
        {Krizhevsky:2012:nips}
\bibfield{author}{\bibinfo{person}{Alex Krizhevsky}, \bibinfo{person}{Ilya
  Sutskever}, {and} \bibinfo{person}{Geoffrey~E Hinton}.}
  \bibinfo{year}{2012}\natexlab{}.
\newblock \showarticletitle{ImageNet Classification with Deep Convolutional
  Neural Networks}.
\newblock In \bibinfo{booktitle}{{\em Advances in Neural Information Processing
  Systems 25}}.
\newblock


\bibitem[\protect\citeauthoryear{LeCun, Bengio, and Hinton}{LeCun
  et~al\mbox{.}}{2015}]%
        {LeCun:2015:nature}
\bibfield{author}{\bibinfo{person}{Yann LeCun}, \bibinfo{person}{Yoshua
  Bengio}, {and} \bibinfo{person}{Geoffrey Hinton}.}
  \bibinfo{year}{2015}\natexlab{}.
\newblock \showarticletitle{Deep Learning}.
\newblock \bibinfo{journal}{{\em Nature\/}} \bibinfo{volume}{521},
  \bibinfo{number}{7553} (\bibinfo{year}{2015}), \bibinfo{pages}{436--444}.
\newblock


\bibitem[\protect\citeauthoryear{Levy and Goldberg}{Levy and Goldberg}{2014}]%
        {Levy:2014:nips}
\bibfield{author}{\bibinfo{person}{Omer Levy} {and} \bibinfo{person}{Yoav
  Goldberg}.} \bibinfo{year}{2014}\natexlab{}.
\newblock \showarticletitle{Neural Word Embedding As Implicit Matrix
  Factorization}. In \bibinfo{booktitle}{{\em Proceedings of the 27th
  International Conference on Neural Information Processing Systems}} {\em
  (\bibinfo{series}{NIPS'14})}.
\newblock


\bibitem[\protect\citeauthoryear{Lowd and Meek}{Lowd and Meek}{2005}]%
        {Lowd:2005:kdd}
\bibfield{author}{\bibinfo{person}{Daniel Lowd} {and}
  \bibinfo{person}{Christopher Meek}.} \bibinfo{year}{2005}\natexlab{}.
\newblock \showarticletitle{Adversarial Learning}. In \bibinfo{booktitle}{{\em
  KDD}}.
\newblock


\bibitem[\protect\citeauthoryear{Marr}{Marr}{2017}]%
        {ml-medical}
\bibfield{author}{\bibinfo{person}{Bernard Marr}.}
  \bibinfo{year}{2017}\natexlab{}.
\newblock \bibinfo{title}{First FDA Approval For Clinical Cloud-Based Deep
  Learning In Healthcare}.
\newblock \bibinfo{howpublished}{\url{https://www.forbes.com/}}.
  (\bibinfo{year}{2017}).
\newblock


\bibitem[\protect\citeauthoryear{{Mikolov}, {Chen}, {Corrado}, and
  {Dean}}{{Mikolov} et~al\mbox{.}}{2013a}]%
        {2013:Mikolov:arxiv}
\bibfield{author}{\bibinfo{person}{T. {Mikolov}}, \bibinfo{person}{K. {Chen}},
  \bibinfo{person}{G. {Corrado}}, {and} \bibinfo{person}{J. {Dean}}.}
  \bibinfo{year}{2013}\natexlab{a}.
\newblock \showarticletitle{{Efficient Estimation of Word Representations in
  Vector Space}}.
\newblock \bibinfo{journal}{{\em ArXiv e-prints\/}} (\bibinfo{year}{2013}).
\newblock


\bibitem[\protect\citeauthoryear{{Mikolov}, {Le}, and {Sutskever}}{{Mikolov}
  et~al\mbox{.}}{2013b}]%
        {2013:Mikolov:arxiv2}
\bibfield{author}{\bibinfo{person}{T. {Mikolov}}, \bibinfo{person}{Q.~V. {Le}},
  {and} \bibinfo{person}{I. {Sutskever}}.} \bibinfo{year}{2013}\natexlab{b}.
\newblock \showarticletitle{{Exploiting Similarities among Languages for
  Machine Translation}}.
\newblock \bibinfo{journal}{{\em ArXiv e-prints\/}} (\bibinfo{year}{2013}).
\newblock


\bibitem[\protect\citeauthoryear{Minka}{Minka}{}]%
        {glossary}
\bibfield{author}{\bibinfo{person}{Thomas Minka}.}
\newblock \bibinfo{title}{A Statistical Learning/Pattern Recognition Glossary}.
\newblock
  \bibinfo{howpublished}{\url{http://alumni.media.mit.edu/~tpminka/statlearn/glossary/}}.
    (\bibinfo{year}{????}).
\newblock


\bibitem[\protect\citeauthoryear{Nelson, Rubinstein, Huang, Joseph, Lee, Rao,
  and Tygar}{Nelson et~al\mbox{.}}{2012}]%
        {Nelson:2012:QSE}
\bibfield{author}{\bibinfo{person}{Blaine Nelson}, \bibinfo{person}{Benjamin
  I.~P. Rubinstein}, \bibinfo{person}{Ling Huang}, \bibinfo{person}{Anthony~D.
  Joseph}, \bibinfo{person}{Steven~J. Lee}, \bibinfo{person}{Satish Rao}, {and}
  \bibinfo{person}{J.~D. Tygar}.} \bibinfo{year}{2012}\natexlab{}.
\newblock \showarticletitle{Query Strategies for Evading Convex-inducing
  Classifiers}.
\newblock \bibinfo{journal}{{\em J. Mach. Learn. Res.\/}}  \bibinfo{volume}{13}
  (\bibinfo{year}{2012}), \bibinfo{pages}{1293--1332}.
\newblock


\bibitem[\protect\citeauthoryear{Papernot, McDaniel, Jha, Fredrikson, Celik,
  and Swamil}{Papernot et~al\mbox{.}}{2016a}]%
        {Papernot:2016:eurosp}
\bibfield{author}{\bibinfo{person}{Nicolas Papernot}, \bibinfo{person}{Patrick
  McDaniel}, \bibinfo{person}{Somesh Jha}, \bibinfo{person}{Matt Fredrikson},
  \bibinfo{person}{Z.~Berkay Celik}, {and} \bibinfo{person}{Ananthram Swamil}.}
  \bibinfo{year}{2016}\natexlab{a}.
\newblock \showarticletitle{The Limitations of Deep Learning in Adversarial
  Settings}. In \bibinfo{booktitle}{{\em Euro S\&P}}.
\newblock


\bibitem[\protect\citeauthoryear{Papernot, McDaniel, Wu, Jha, and
  Swami}{Papernot et~al\mbox{.}}{2016b}]%
        {Papernot:2016:sp}
\bibfield{author}{\bibinfo{person}{Nicolas Papernot}, \bibinfo{person}{Patrick
  McDaniel}, \bibinfo{person}{Xi Wu}, \bibinfo{person}{Somesh Jha}, {and}
  \bibinfo{person}{Ananthram Swami}.} \bibinfo{year}{2016}\natexlab{b}.
\newblock \showarticletitle{Distillation as a Defense to Adversarial
  Perturbations against Deep Neural Networks}. In \bibinfo{booktitle}{{\em
  S\&P}}.
\newblock


\bibitem[\protect\citeauthoryear{{Park} and {Boyd}}{{Park} and {Boyd}}{2017}]%
        {Park:2017:arxiv}
\bibfield{author}{\bibinfo{person}{J. {Park}} {and} \bibinfo{person}{S.
  {Boyd}}.} \bibinfo{year}{2017}\natexlab{}.
\newblock \showarticletitle{{General Heuristics for Nonconvex Quadratically
  Constrained Quadratic Programming}}.
\newblock \bibinfo{journal}{{\em ArXiv e-prints\/}} (\bibinfo{year}{2017}).
\newblock


\bibitem[\protect\citeauthoryear{Pennington, Socher, and Manning}{Pennington
  et~al\mbox{.}}{2014}]%
        {Pennington:2014:emnlp}
\bibfield{author}{\bibinfo{person}{Jeffrey Pennington},
  \bibinfo{person}{Richard Socher}, {and} \bibinfo{person}{Christopher~D.
  Manning}.} \bibinfo{year}{2014}\natexlab{}.
\newblock \showarticletitle{GloVe: Global Vectors for Word Representation}.
  \bibinfo{howpublished}{\url{http://github.com/stanfordnlp/glove}}. In
  \bibinfo{booktitle}{{\em Empirical Methods in Natural Language Processing
  (EMNLP)}}.
\newblock


\bibitem[\protect\citeauthoryear{Roesner and Kohno}{Roesner and Kohno}{2013}]%
        {Roesner:2013:sec}
\bibfield{author}{\bibinfo{person}{Franziska Roesner} {and}
  \bibinfo{person}{Tadayoshi Kohno}.} \bibinfo{year}{2013}\natexlab{}.
\newblock \showarticletitle{Securing Embedded User Interfaces: Android and
  Beyond}. In \bibinfo{booktitle}{{\em Proceedings of the 22Nd USENIX
  Conference on Security}} {\em (\bibinfo{series}{SEC'13})}.
\newblock


\bibitem[\protect\citeauthoryear{Rubinstein, Nelson, Huang, Joseph, Lau, Rao,
  Taft, and Tygar}{Rubinstein et~al\mbox{.}}{2009}]%
        {Rubinstein:2009:imc}
\bibfield{author}{\bibinfo{person}{Benjamin~I.P. Rubinstein},
  \bibinfo{person}{Blaine Nelson}, \bibinfo{person}{Ling Huang},
  \bibinfo{person}{Anthony~D. Joseph}, \bibinfo{person}{Shing-hon Lau},
  \bibinfo{person}{Satish Rao}, \bibinfo{person}{Nina Taft}, {and}
  \bibinfo{person}{J.~D. Tygar}.} \bibinfo{year}{2009}\natexlab{}.
\newblock \showarticletitle{ANTIDOTE: Understanding and Defending Against
  Poisoning of Anomaly Detectors}. In \bibinfo{booktitle}{{\em IMC}}.
\newblock


\bibitem[\protect\citeauthoryear{Russakovsky, Deng, Su, Krause, Satheesh, Ma,
  Huang, Karpathy, Khosla, Bernstein, Berg, and Fei-Fei}{Russakovsky
  et~al\mbox{.}}{2015}]%
        {ILSVRC15}
\bibfield{author}{\bibinfo{person}{Olga Russakovsky}, \bibinfo{person}{Jia
  Deng}, \bibinfo{person}{Hao Su}, \bibinfo{person}{Jonathan Krause},
  \bibinfo{person}{Sanjeev Satheesh}, \bibinfo{person}{Sean Ma},
  \bibinfo{person}{Zhiheng Huang}, \bibinfo{person}{Andrej Karpathy},
  \bibinfo{person}{Aditya Khosla}, \bibinfo{person}{Michael Bernstein},
  \bibinfo{person}{Alexander~C. Berg}, {and} \bibinfo{person}{Li Fei-Fei}.}
  \bibinfo{year}{2015}\natexlab{}.
\newblock \showarticletitle{{ImageNet Large Scale Visual Recognition
  Challenge}}.
\newblock \bibinfo{journal}{{\em International Journal of Computer Vision
  (IJCV)\/}} \bibinfo{volume}{115}, \bibinfo{number}{3} (\bibinfo{year}{2015}),
  \bibinfo{pages}{211--252}.
\newblock


\bibitem[\protect\citeauthoryear{Satariano}{Satariano}{2017}]%
        {ml-financial}
\bibfield{author}{\bibinfo{person}{Adam Satariano}.}
  \bibinfo{year}{2017}\natexlab{}.
\newblock \bibinfo{title}{AI trader? Tech vet launches hedge fund run by
  artificial intelligence}.
\newblock \bibinfo{howpublished}{\url{http://www.dailyherald.com/}}.
  (\bibinfo{year}{2017}).
\newblock


\bibitem[\protect\citeauthoryear{Sculley, Holt, Golovin, Davydov, Phillips,
  Ebner, Chaudhary, Young, Crespo, and Dennison}{Sculley et~al\mbox{.}}{2015}]%
        {Sculley:2015:nips}
\bibfield{author}{\bibinfo{person}{D. Sculley}, \bibinfo{person}{Gary Holt},
  \bibinfo{person}{Daniel Golovin}, \bibinfo{person}{Eugene Davydov},
  \bibinfo{person}{Todd Phillips}, \bibinfo{person}{Dietmar Ebner},
  \bibinfo{person}{Vinay Chaudhary}, \bibinfo{person}{Michael Young},
  \bibinfo{person}{Jean-Francois Crespo}, {and} \bibinfo{person}{Dan
  Dennison}.} \bibinfo{year}{2015}\natexlab{}.
\newblock \showarticletitle{Hidden Technical Debt in Machine Learning Systems}.
  In \bibinfo{booktitle}{{\em Proceedings of the 28th International Conference
  on Neural Information Processing Systems}} {\em (\bibinfo{series}{NIPS'15})}.
\newblock


\bibitem[\protect\citeauthoryear{Security and Crash Bugs of Libpng}{Security
  and Crash Bugs of Libpng}{2004}]%
        {libpng}
Security and Crash Bugs of Libpng \bibinfo{year}{2004}\natexlab{}.
\newblock
  \bibinfo{howpublished}{\url{http://www.libpng.org/pub/png/libpng.html}}.
  (\bibinfo{year}{2004}).
\newblock


\bibitem[\protect\citeauthoryear{Sharif, Bhagavatula, Bauer, and Reiter}{Sharif
  et~al\mbox{.}}{2016}]%
        {Sharif:2016:ccs}
\bibfield{author}{\bibinfo{person}{Mahmood Sharif}, \bibinfo{person}{Sruti
  Bhagavatula}, \bibinfo{person}{Lujo Bauer}, {and} \bibinfo{person}{Michael~K.
  Reiter}.} \bibinfo{year}{2016}\natexlab{}.
\newblock \showarticletitle{Accessorize to a Crime: Real and Stealthy Attacks
  on State-of-the-Art Face Recognition}. In \bibinfo{booktitle}{{\em
  Proceedings of the 2016 ACM SIGSAC Conference on Computer and Communications
  Security}} {\em (\bibinfo{series}{CCS '16})}.
\newblock


\bibitem[\protect\citeauthoryear{{Simonyan} and {Zisserman}}{{Simonyan} and
  {Zisserman}}{2014}]%
        {Simonyan:2014:arxiv}
\bibfield{author}{\bibinfo{person}{K. {Simonyan}} {and} \bibinfo{person}{A.
  {Zisserman}}.} \bibinfo{year}{2014}\natexlab{}.
\newblock \showarticletitle{{Very Deep Convolutional Networks for Large-Scale
  Image Recognition}}.
\newblock \bibinfo{journal}{{\em ArXiv e-prints\/}} (\bibinfo{year}{2014}).
\newblock


\bibitem[\protect\citeauthoryear{Szegedy, Liu, Jia, Sermanet, Reed, Anguelov,
  Erhan, Vanhoucke, and Rabinovich}{Szegedy et~al\mbox{.}}{2015}]%
        {Szegedy:2015:cvpr}
\bibfield{author}{\bibinfo{person}{C. Szegedy}, \bibinfo{person}{Wei Liu},
  \bibinfo{person}{Yangqing Jia}, \bibinfo{person}{P. Sermanet},
  \bibinfo{person}{S. Reed}, \bibinfo{person}{D. Anguelov}, \bibinfo{person}{D.
  Erhan}, \bibinfo{person}{V. Vanhoucke}, {and} \bibinfo{person}{A.
  Rabinovich}.} \bibinfo{year}{2015}\natexlab{}.
\newblock \showarticletitle{Going deeper with convolutions}. In
  \bibinfo{booktitle}{{\em 2015 IEEE Conference on Computer Vision and Pattern
  Recognition (CVPR)}}.
\newblock


\bibitem[\protect\citeauthoryear{{Szegedy}, {Vanhoucke}, {Ioffe}, {Shlens}, and
  {Wojna}}{{Szegedy} et~al\mbox{.}}{2015}]%
        {Szegedy:2015:arxiv}
\bibfield{author}{\bibinfo{person}{C. {Szegedy}}, \bibinfo{person}{V.
  {Vanhoucke}}, \bibinfo{person}{S. {Ioffe}}, \bibinfo{person}{J. {Shlens}},
  {and} \bibinfo{person}{Z. {Wojna}}.} \bibinfo{year}{2015}\natexlab{}.
\newblock \showarticletitle{{Rethinking the Inception Architecture for Computer
  Vision}}.
\newblock \bibinfo{journal}{{\em ArXiv e-prints\/}} (\bibinfo{year}{2015}).
\newblock


\bibitem[\protect\citeauthoryear{{Szegedy}, {Zaremba}, {Sutskever}, {Bruna},
  {Erhan}, {Goodfellow}, and {Fergus}}{{Szegedy} et~al\mbox{.}}{2013}]%
        {Szegedy:2013:arxiv}
\bibfield{author}{\bibinfo{person}{C. {Szegedy}}, \bibinfo{person}{W.
  {Zaremba}}, \bibinfo{person}{I. {Sutskever}}, \bibinfo{person}{J. {Bruna}},
  \bibinfo{person}{D. {Erhan}}, \bibinfo{person}{I. {Goodfellow}}, {and}
  \bibinfo{person}{R. {Fergus}}.} \bibinfo{year}{2013}\natexlab{}.
\newblock \showarticletitle{{Intriguing properties of neural networks}}.
\newblock \bibinfo{journal}{{\em ArXiv e-prints\/}} (\bibinfo{year}{2013}).
\newblock


\bibitem[\protect\citeauthoryear{{Tabacof} and {Valle}}{{Tabacof} and
  {Valle}}{2015}]%
        {Tabacof:2015:arXiv}
\bibfield{author}{\bibinfo{person}{P. {Tabacof}} {and} \bibinfo{person}{E.
  {Valle}}.} \bibinfo{year}{2015}\natexlab{}.
\newblock \showarticletitle{{Exploring the Space of Adversarial Images}}.
\newblock \bibinfo{journal}{{\em ArXiv e-prints\/}} (\bibinfo{year}{2015}).
\newblock


\bibitem[\protect\citeauthoryear{The Heartbleed Bug}{The Heartbleed
  Bug}{2014}]%
        {heartbleed}
The Heartbleed Bug \bibinfo{year}{2014}\natexlab{}.
\newblock \bibinfo{howpublished}{\url{http://heartbleed.com}}.
  (\bibinfo{year}{2014}).
\newblock


\bibitem[\protect\citeauthoryear{Tram{\`e}r, Zhang, Juels, Reiter, and
  Ristenpart}{Tram{\`e}r et~al\mbox{.}}{2016}]%
        {Tramer:2016:sec}
\bibfield{author}{\bibinfo{person}{Florian Tram{\`e}r}, \bibinfo{person}{Fan
  Zhang}, \bibinfo{person}{Ari Juels}, \bibinfo{person}{Michael~K. Reiter},
  {and} \bibinfo{person}{Thomas Ristenpart}.} \bibinfo{year}{2016}\natexlab{}.
\newblock \showarticletitle{Stealing Machine Learning Models via Prediction
  APIs}. In \bibinfo{booktitle}{{\em 25th USENIX Security Symposium (USENIX
  Security 16)}}.
\newblock


\bibitem[\protect\citeauthoryear{Veracode}{Veracode}{2014}]%
        {veracode}
\bibfield{author}{\bibinfo{person}{Veracode}.} \bibinfo{year}{2014}\natexlab{}.
\newblock \bibinfo{title}{Open Source and Third-Party Components Embed 24 Known
  Vulnerabilities into Every Web Application on Average}.
\newblock \bibinfo{howpublished}{\url{https://www.veracode.com/}}.
  (\bibinfo{year}{2014}).
\newblock


\bibitem[\protect\citeauthoryear{Xiao, Biggio, Nelson, Xiao, Eckert, and
  Roli}{Xiao et~al\mbox{.}}{2015}]%
        {Xiao:2015:SVM}
\bibfield{author}{\bibinfo{person}{Huang Xiao}, \bibinfo{person}{Battista
  Biggio}, \bibinfo{person}{Blaine Nelson}, \bibinfo{person}{Han Xiao},
  \bibinfo{person}{Claudia Eckert}, {and} \bibinfo{person}{Fabio Roli}.}
  \bibinfo{year}{2015}\natexlab{}.
\newblock \showarticletitle{Support Vector Machines Under Adversarial Label
  Contamination}.
\newblock \bibinfo{journal}{{\em Neurocomput.\/}} \bibinfo{volume}{160},
  \bibinfo{number}{C} (\bibinfo{year}{2015}), \bibinfo{pages}{53--62}.
\newblock


\bibitem[\protect\citeauthoryear{{Yosinski}, {Clune}, {Bengio}, and
  {Lipson}}{{Yosinski} et~al\mbox{.}}{2014}]%
        {Yosinski:2014:nips}
\bibfield{author}{\bibinfo{person}{J. {Yosinski}}, \bibinfo{person}{J.
  {Clune}}, \bibinfo{person}{Y. {Bengio}}, {and} \bibinfo{person}{H.
  {Lipson}}.} \bibinfo{year}{2014}\natexlab{}.
\newblock \showarticletitle{{How transferable are features in deep neural
  networks?}}
\newblock \bibinfo{journal}{{\em ArXiv e-prints\/}} (\bibinfo{year}{2014}).
\newblock


\bibitem[\protect\citeauthoryear{{Zhang}, {Bengio}, {Hardt}, {Recht}, and
  {Vinyals}}{{Zhang} et~al\mbox{.}}{2016}]%
        {Zhang:2016:arxiv}
\bibfield{author}{\bibinfo{person}{C. {Zhang}}, \bibinfo{person}{S. {Bengio}},
  \bibinfo{person}{M. {Hardt}}, \bibinfo{person}{B. {Recht}}, {and}
  \bibinfo{person}{O. {Vinyals}}.} \bibinfo{year}{2016}\natexlab{}.
\newblock \showarticletitle{{Understanding deep learning requires rethinking
  generalization}}.
\newblock \bibinfo{journal}{{\em ArXiv e-prints\/}} (\bibinfo{year}{2016}).
\newblock


\end{thebibliography}

\section*{Appendix}

\subsection*{Multi-Target Logic-Bomb Attacks}
\label{sec:group}

 We now generalize the single target attacks to the case of multiple targets $\mathcal{D} = \{(\vxs, \vys)\}$.

%
%

 A straightforward solution is to sequentially apply Algorithm~\ref{alg:bomb} on each target of $\mathcal{D}$. This solution however suffers the drawback that both the number of perturbed parameters and the impact on the classification of non-target inputs accumulate with the number of targets.

 We overcome this limitation by introducing the definition of multi-target positive impact:
 $|\sum_{(\vxs, \vys) \in \mathcal{T}}\Delta_{(\vxs, \vys)}^\theta|$,
  which quantifies $\theta$'s overall influence on these targets. It is worth noting the difference between the definitions of positive and negative impact (i.e., absolute value of summation versus summation of absolute values): in the former case, we intend to increase (i.e., directional) the likelihood of target inputs being classified into desired classes; in the latter case, we intend to minimize the impact (i.e., directionless) on the classification of all non-target inputs.

 By substituting the single-target positive impact measure with its multi-target version, Algorithm~\ref{alg:bomb2} can be readily generalized to craft \mlcs targeting multiple inputs. We omit the details here.

\end{document}